\documentclass[acmtog]{acmart}
\acmSubmissionID{xxx}
\AtBeginDocument{%
  }

\copyrightyear{2025}
\acmYear{2025}
\acmConference[SIGGRAPH Conference Papers '25]{Special Interest Group on Computer Graphics and Interactive Techniques Conference Conference Papers }{August 10--14, 2025}{Vancouver, BC, Canada}
\acmBooktitle{Special Interest Group on Computer Graphics and Interactive Techniques Conference Conference Papers (SIGGRAPH Conference Papers '25), August 10--14, 2025, Vancouver, BC, Canada}
\acmDOI{10.1145/3721238.3730635}
\acmISBN{979-8-4007-1540-2/2025/08}

\usepackage{graphicx}
\usepackage{bbding}

\usepackage{booktabs}
\usepackage{ulem}

\usepackage{multirow}
\usepackage{amsmath,bm}
\usepackage{url}
\usepackage[ruled]{algorithm2e} 

\SetAlFnt{\small}
\SetAlCapFnt{\small}
\SetAlCapNameFnt{\small}
\SetAlCapHSkip{0pt}
\usepackage{overpic} 
\usepackage{subcaption}
\usepackage{array}
\usepackage{float}

\usepackage{xcolor} 
\usepackage{colortbl}

\definecolor{shadecolor}{rgb}{0.92, 0.92, 0.92}
\definecolor{gtgray}{gray}{0.97}
\definecolor{mygray}{gray}{.88}

\definecolor{gray1}{gray}{.90}
\definecolor{gray2}{gray}{.92}
\definecolor{gray3}{gray}{.94}

\makeatletter
\def\hlinew#1{%
  \noalign{\ifnum0=`}\fi\hrule \@height #1 \futurelet
   \reserved@a\@xhline}
\makeatother
\usepackage[capitalize]{cleveref}
\crefname{section}{Sec.}{Secs.}
\Crefname{section}{Section}{Sections}
\Crefname{table}{Table}{Tables}
\crefname{table}{Tab.}{Tabs.}

\usepackage{array}
\newcolumntype{L}[1]{>{\raggedright\let\newline\\\arraybackslash\hspace{0pt}}m{#1}}
\newcolumntype{C}[1]{>{\centering\let\newline\\\arraybackslash\hspace{0pt}}m{#1}}
\newcolumntype{R}[1]{>{\raggedleft\let\newline\\\arraybackslash\hspace{0pt}}m{#1}}

\long\def\ignorethis#1{}
\definecolor{crimson}{rgb}{0.86, 0.08, 0.24}
\definecolor{green}{rgb}{0, 0.5, 0.25}
\definecolor{purple}{rgb}{0.75, 0, 1}
\definecolor{orange}{rgb}{1, 0.5, 0.25}
\definecolor{yellow}{rgb}{1, 1, 0}
\definecolor{new_blue}{rgb}{0, 0.5, 1}
\definecolor{new_cyan}{rgb}{0.10, 0.62, 0.57}

\begin{document}

\title{StreamME: Simplify 3D Gaussian Avatar within Live Stream}

\author{Luchuan Song}
\email{lsong11@cs.rochester.edu}
\affiliation{%
  \institution{University of Rochester}
  \country{USA}
}
\author{Yang Zhou}
\affiliation{%
  \institution{Adobe Research}
  \country{USA}
}

\author{Zhan Xu}
\affiliation{%
  \institution{Adobe Research}
  \country{USA}
}

\author{Yi Zhou}
\affiliation{%
  \institution{Adobe Research}
  \country{USA}
}

\author{Deepali Aneja}
\affiliation{%
  \institution{Adobe Research}
  \country{USA}
}

\author{Chenliang Xu}
\affiliation{%
  \institution{University of Rochester}
  \country{USA}
}

\begin{abstract}
We propose StreamME, a method focuses on fast 3D avatar reconstruction. The StreamME synchronously records and reconstructs a head avatar from live video streams without any pre-cached data, enabling seamless integration of the reconstructed appearance into downstream applications. This exceptionally fast training strategy, which we refer to as on-the-fly training, is central to our approach. Our method is built upon 3D Gaussian Splatting (3DGS), eliminating the reliance on MLPs in deformable 3DGS and relying solely on geometry, which significantly improves the adaptation speed to facial expression. To further ensure high efficiency in on-the-fly training, we introduced a simplification strategy based on primary points, which distributes the point clouds more sparsely across the facial surface, optimizing points number while maintaining rendering quality. Leveraging the on-the-fly training capabilities, our method protects the facial privacy and reduces communication bandwidth in VR system or online conference. Additionally, it can be directly applied to downstream application such as animation, toonify, and relighting. Please refer to our project page for more details: \url{https://songluchuan.github.io/StreamME/}.
\end{abstract}

%


\begin{teaserfigure}
\includegraphics[width=1.\textwidth]{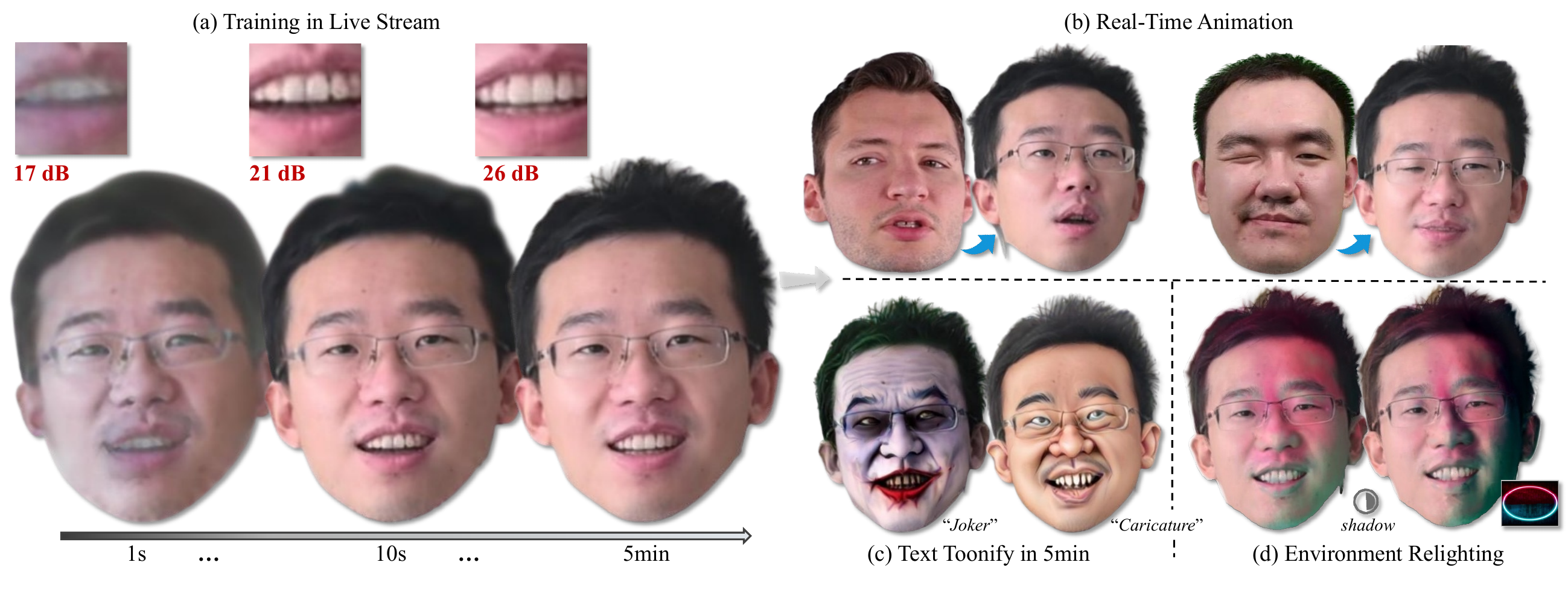}
\begin{small}
\end{small}
\vspace{-0.9cm}
\caption{The \textbf{StreamME} takes live stream (or monocular) video as input to enable rapid 3D head avatar reconstruction. It achieves impressive speed, capturing the basic facial appearance within 10 seconds (PSNR = 21 dB) and reaching high-quality fidelity (PSNR = 26 dB) within 5 minutes, as shown in (a). Notably, StreamME reconstructs facial features through on-the-fly training, allowing simultaneous recording and modeling without the need for pre-cached data (\textit{e.g.}~pre-training model). Additionally, StreamME facilitates real-time animation in (b), toonify in (c) and relighting in (d) (the background light image is shown in bottom right and we move the light position to create shadows on face) from the 5-minute reconstructed appearance, supporting the applications in VR and online conference. Natural face$\copyright$\textit{Xuan Gao et al.} (CC BY), and $\copyright$\textit{Wojciech Zielonka et al.} (CC BY).}
\label{fig:teaser}
\end{teaserfigure}

\maketitle

\section{Introduction}\label{sec:intro}

The rapid reconstruction of head avatars reconstruction and reenactment of facial expression dynamics from a single video have become a rapidly advancing research tpoic, with vast potential for applications in VR/AR, digital human development, holographic communication, live streaming, and more. Recently, the volumetric models (\textit{e.g.}~Instance-NeRF~\cite{liu2023instance} and 3DGS~\cite{kerbl20233d}) have endeavored to achieve both high-quality and efficient rendering. 
For instance, 
INSTA~\cite{zielonka2022towards} employs Instant-NGP~\cite{muller2022instant} to accelerate rendering through engineering optimizations.
AvatarMAV~\cite{xu2023avatarmav} leverages the learnable blendshape as motion representation to achieve fast recovery of head avatar.  FlashAvatar~\cite{xiang2023flashavatar} simulates the head avatar with a large number of Gaussian points in UV space. 
However, they continue to face challenges in balancing rendering quality and storage overhead, which constrains their applicability in consumer applications.
\par

In this paper, we advance rapid facial reconstruction techniques to address the existing limitations. Additionally, we introduce a novel head avatar reconstruction task, termed on-the-fly training for reconstruction, which pushes the efficiency boundaries of fast reconstruction even further. Based on these observations, prior methods have uniformly separated training and inference processes due to efficiency constraints (refer as offline training). Such as, while AvatarMAV~\cite{xu2023avatarmav} achieves efficient training speeds offline, it cannot support frame-by-frame training for reconstruction within the live streaming. Our on-the-fly training approach offers multiple advantages, including (i) protect facial privacy by eliminating the need to pre-cached personal facial models on the external machines, (ii) only 3DGS parameters are transmitted in stream video, rather than the full images (about 70\% compression) and (iii) the synchronous training and recording with real-time visualization, allowing for immediate re-recording of under-trained facial areas. \par

We propose a novel on-the-fly head avatar reconstruction method named \textbf{StreamME}. Different from the all previous head avatar reconstruction methods~\cite{xiang2023flashavatar,xu2023avatarmav,qian2023gaussianavatars,song2021tacr,xu2023gaussianheadavatar,wang2023styleavatar,zheng2022avatar,zheng2023pointavatar}, the StreamME avoids dependence on multiple MLP layers to capture deformable facial dynamics (\textit{e.g.}, facial expression motion), significantly reducing expression recovery time and enabling true on-the-fly training. Specifically, we attach the 3D Gaussian point clouds to the tracked head mesh surface, allowing the points to move in tandem with mesh deformations. However, the point clouds associated with the deformed mesh on the 3D head template do not fully preserve the geometric properties of the face, which results in noise cloud artifacts around the rendered face and reducing the realism. From our method, we dynamically adjust the initial 3D Gaussian points through anchor-based pruning-and-clone strategy. Instead of selecting all points from the tracked head mesh as 3D Gaussian points, we identify specific anchor points that accurately capture facial motion. The 3D Gaussian points are then updated based on these selected anchors, optimizing for head representation. This strategy improves efficiency from eliminating points that do not contribute to facial motion, while preserving the motion anchor points critical for controlling facial deformation. \par 

Meanwhile, we find in practice that more replicated 3D Gaussian points will lead to better quality but reduce speed, especially the training speed involving backpropagation. Therefore, we explored a method to gradually simplify the point clouds, which balance the number of point clouds and rendering quality. Here, we introduce two assumptions for simplifying point clouds: (i) the points should be distributed around the facial surface, rather than within it, as internal points remain unobservable due to occlusion; (ii) the small-size 3D Gaussian points with minimal volume, contribute negligibly to image quality and the impact is imperceptible. In optimization, these two assumptions serve as foundational principles. We ensure that 3D Gaussians are progressively distributed around and outside the surface, while occluded and small-sized points are removed, enhancing execution speed. This strategy yields a sparser 3D Gaussian representation of the head avatar, substantially improving efficiency without compromising rendering quality.\par

With the help of \textit{Motion-Aware Anchor Points} selection and \textit{Gaussian Points Simplification} strategy, we achieve on-the-fly photo-realistic head avatar representation within approximately 5 minutes of live streaming, as shown in Figure~\ref{fig:teaser} (a). Moreover, the 3D Gaussian properties learned within 5 minutes can be applied to cross-identity head animation, facial toonification, environment relighting, and other applications with minimal fine-tuning, as illustrated in Figure~\ref{fig:teaser} (b, c, d). This flexibility significantly broadens the application scope of our method. Furthermore, we demonstrate the superiority of our method through extensive experiments and comparisons with both instant and long-term training approaches. In summary, our contributions include the following aspects: \par 

\noindent \textbf{(1)} We present the on-the-fly head avatar reconstruction method, which is able to reconstruct facial appearance from the live streams within about 5 minutes by pure Pytorch code. To the best of our knowledge, we are the first to reconstruct and visualize the head avatar within the on-the-fly training. \par 

\noindent \textbf{(2)} We emphasize the efficiency in training, and introduce motion saliency anchor selection and point cloud simplification strategy. The anchor selection minimizes reliance on MLPs within the deformation field, while point cloud simplification strategy reduces computational redundancy from 3D Gaussian points.

\noindent \textbf{(3)} A series of downstream applications are attached, which have demonstrated the advances of our approach and provided novel insight for the on-the-fly training method.

\begin{figure*}[t]
  \centering
  \includegraphics[width=.99\linewidth]{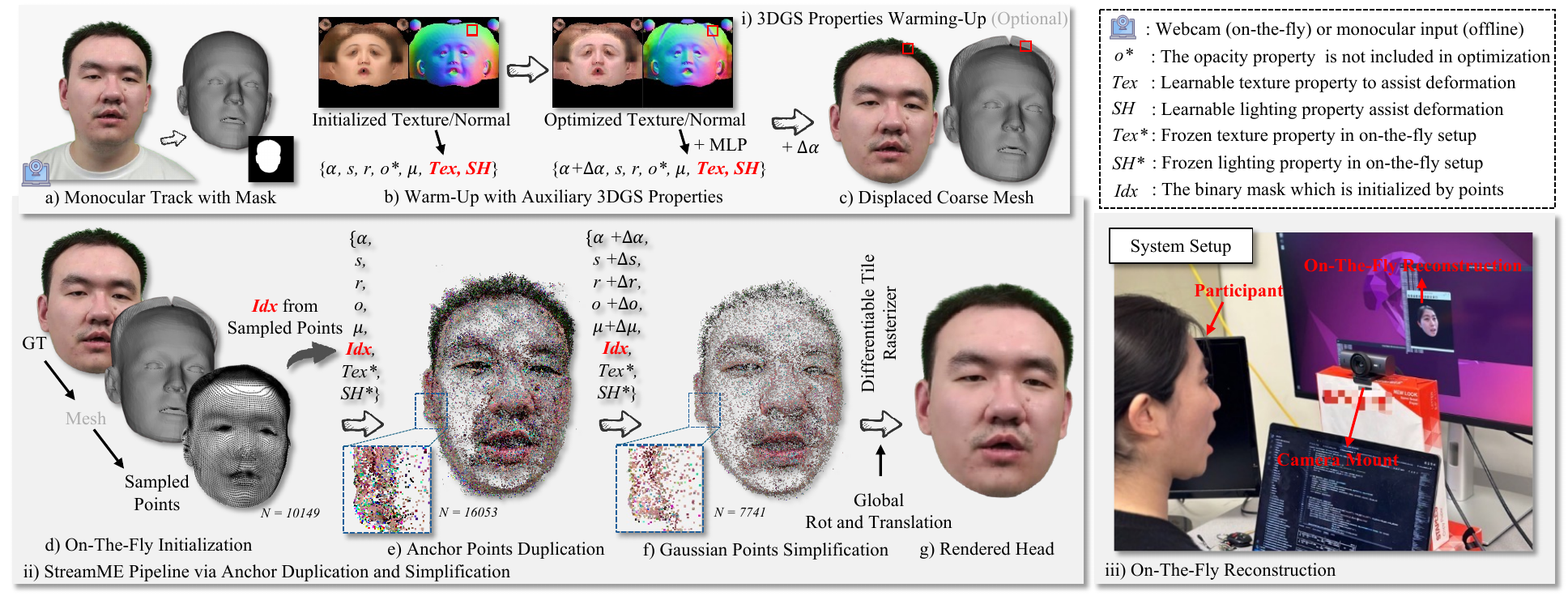}
  \vspace{-0.4cm}
  \caption{The overview the pipeline of \textbf{StreamME}. We list three components at here. i) The 3DGS Properties Warming-Up (Optional): we introduce two auxiliary learnable 3D Gaussian attribute texture and illumination, refining the UV vertex positions to improve facial geometry detail (\textit{e.g.} here, we show the coarse displacement for the vertices around the hair). This step is optional, and the users may also opt to use the tracked head without displacement. ii) Anchor Duplication and Simplification: we freeze the Tex and SH attributes and introduce a binary learnable mask, initialized with all values set to $1$, from the UV vertices sampled on the mesh. Natural face$\copyright$\textit{Xuan Gao et al.} (CC BY).}
  \vspace{-0.1cm}
  \label{fig_pipeline}
\end{figure*}

%

\section{Related Works}

\subsection{Instant Head Avatar}

The instant head avatar is an evolving field rooted in traditional photorealistic head reconstruction, and improved by novel techniques that reduce dependency on long-term training (\textit{e.g.}, StyleAvatar~\cite{wang2023styleavatar}, IM Avatar~\cite{zheng2022avatar}, PointAvatar~\cite{zheng2023pointavatar}, Deep-Video-Portrait~\cite{kim2018deep}, NeRFace~\cite{Gafni_2021_CVPR}, KeypointNeRF~\cite{mihajlovic2022keypointnerf}~\textit{e.t.c}) to achieve high-quality results. Recently, with the introduction of 3DGS~\cite{kerbl20233d} and NeRF acceleration~\cite{muller2022instant} has driven rapid advancements in this field. The most notable works in this area include INSTA~\cite{zielonka2023instant}, AvatarMAV~\cite{xu2023avatarmav}, and FlashAvatar~\cite{xiang2023flashavatar}. And the INSTA and FlashAvatar utilize mesh geometry sampling, with Instant-NGP and 3DGS employed for accelerated rendering, respectively. The AvatarMAV~\cite{xu2023avatarmav} employs blendshapes and uses the learnable MLPs to blend multiple implicit representations. Beyond these approaches, other head avatar reconstruction methods~\cite{wang2023styleavatar,zheng2023pointavatar,Gafni_2021_CVPR,kim2018deep,thies2016face2face} usually require several hours to several days to complete. \par

Additionally, the above methods utilize linearly decomposed facial expression parameters as coarse conditions, by neural convolution or MLP layers to refine the details such as hair and mouth details. In contrast to these approaches, we maintain high fidelity with more efficient in training and inference. With only the 3D Gaussian primitives attached to explicit geometry, we significantly reduce the learning burden from learnable neural networks and accelerate the training process. To our knowledge, this is the first work to achieve on-the-fly training for head avatar reconstruction.

\subsection{Real-Time Face Estimation}

The on-the-fly head avatar reconstruction depends on both efficient 3DGS training and real-time 3D head geometry estimation. As in previous works~\cite{wang2022hvh,xiang2023flashavatar,xu2023avatarmav,shao2024splattingavatar,song2021fsft,liu2022deepfacevideoediting,chan2022efficient,cao2022authentic,song2023emotional}, the preprocessed data is derived from estimating the pose and deformation of the facial template with tools such as MICA~\cite{MICA:ECCV2022}. It is highly time-consuming, for instance, processing a 10-minute video can require one day, which severely limiting its real-time applicability. In our case, we integrate a real-time face estimation module to supply pose and head mesh data on-the-fly within the pipeline. Notably, we have meticulously designed the system to balance resource allocation across facial parsing, head tracking, and 3DGS reconstruction within the real-time pipeline.
%

\subsection{Deformable-3DGS Representations}
The 3D Gaussian Splatting (3DGS)~\cite{kerbl20233d} is designed to represent and render static 3D scenes. Building on static 3DGS, deformable-3DGS~\cite{qian20243dgs,bae2024per,zhu2024motiongs,chen2024monogaussianavatar,tu2024tele,luiten2023dynamic,li2023spacetime,song2024tri} incorporates MLPs or CNNs to predict and render geometric deformations, with 3D Gaussian properties predicted frame-by-frame over time. Specifically, these methods retain a canonical 3D Gaussian space, optimizing the MLP-based deformation field~\cite{zhang2024bag,zhang2023deformtoon3d,zhang2024discover} conditioned on timestamps. \par

Our method also employs the canonical-to-world space strategy. First, the tracked deformed meshes are positioned within canonical space. Then, 3D Gaussian points are placed around the surface and transformed into world space, incorporating pose information for rendering. Additionally, our method eliminates the need for learnable layers to predict deformations, instead, the deformations are derived from 3D Gaussian points around the meshes which are deformed from canonical space. \par

\begin{figure*}[t]
  \centering
  \includegraphics[width=.99\linewidth]{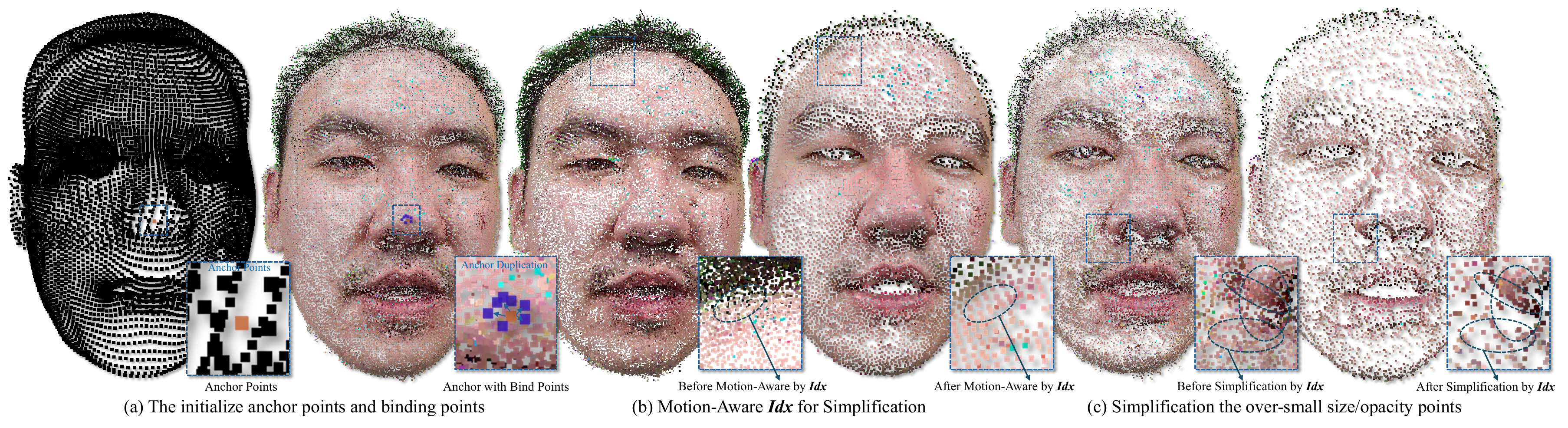}
  \caption{We employ the proposed \textit{\textbf{Idx}} for anchor binding, elimination the motion irrelevant points and simplification the over-small size/opacity points to reduce computational overhead. Specifically, (a) The anchor points are sampled from the mesh and multiple duplicated for detail representation. (b) The motion-aware \textit{\textbf{Idx}} tends to remove points, as there are none motion gradients around the forehead in canonical space. (c) The learnable masks from \textit{\textbf{Idx}} for deleting the small size and opacity points within training. Please zoom-in for details. Natural face$\copyright$\textit{Xuan Gao et al.} (CC BY).
  }
  \label{fig_simpliciation}
\end{figure*}

\section{Method}

The StreamME generates the 3D head avatar within few minutes from streaming. And progressively simplifying the 3D Gaussian point representation of the face while retaining anchor points essential to capturing facial motion. In this section, we will introduce the simplification 3D Gaussians representation and the system pipeline, which is also shown in Figure~\ref{fig_pipeline}. \par

\subsection{Preliminaries}
In our approach, we build upon the 3DGS~\cite{kerbl20233d}, a point-based representation from 3D Gaussian properties for dynamic 3D head reconstruction. In 3DGS, a covariance matrix ($\Sigma$) and the mean of point ($\mu$) are defined for each Gaussian $g$: 
\begin{equation}
g = e^{-\frac{1}{2}(\mathbf{x} -\mu)^T \Sigma^{-1} (\mathbf{x} -\mu)}.
\label{3dgs}
\end{equation}
For differentiable optimization, the covariance matrix $\Sigma$ is decomposed into a rotation matrix $\mathbf{R}$ and a scaling matrix $\mathbf{S}$, parameterized by a learnable quaternion $r$ and scaling vector $s$. 
For spatial attributes, the 3D Gaussians possess the color of the projected pixels are from the splatting and overlapping of the 3D Gaussians points: 
\begin{equation}
\mathbf{C} = \sum_{i \in N} \mathbf{c}_i \alpha_i \prod_{j=1}^{N-1} (1-\alpha_j).
\label{rs}
\end{equation}
Here, $\alpha_i$ represents the density derived from the projection of 3D Gaussians with $\Sigma$ and opacity $o$, and $N$ denotes the number of points. Consequently, we define the 3D Gaussian attributes as follows:
\begin{equation}
g = \{\alpha, s, r, o, \mu\}.
\label{rs}
\end{equation}
The attributes are optimized in backpropagation rendering pipeline. Generally, Gaussian properties derived from precise geometry result in higher rendering quality.




\subsection{Gaussian Properties Warm-up}
Inspired by previous works~\cite{qian2023gaussianavatars,xu2023gaussianheadavatar,zielonka2024gaussian}, we adjust vertices in the tracked coarse head geometry to account for deformations caused by features like hair, the standard facial templates fall short in representing the full head shape. However, the preprocessing step in those method (\textit{e.g.} DMTet~\cite{shen2021deep} or VHAP~\cite{qian2024versatile}) require substantial time to reorient the geometry and frequently suffer from collapse due to the limitations of single or sparse views. \par 

We propose a warm-up step that introduces additional learnable texture (\textit{\textbf{Tex}}) and illumination (\textit{\textbf{SH}}) parameters to assist vertex deformation via 3DGS, achieving reorientation within 20 seconds. Specifically, we follow FlashAvatar~\cite{xiang2023flashavatar} to initialize the 3D Gaussian points from UV coordinates. Then, the vertex positions are learned in 3D Gaussian properties $\alpha$. We apply the learnable texture with illumination to the diffuse color features of each 3D Gaussian point and the normal consistency between adjacent faces on a mesh surface to keep the smoothness, as shown in Figure~\ref{fig_pipeline}. It is worth noting that this explicit geometric deformation step is optional and could be skipped to improve the operation efficiency. Furthermore, the auxiliary learnable parameters $\textit{\textbf{Tex}}$ and $\textit{\textbf{SH}}$ are optimized only during the warm-up step and remain frozen in subsequent stages. In practice, we set the static Direct-Current component (the self.feature\_dc) in 3DGS as the $\textit{\textbf{SH}}$.\par

\subsection{Motion-Aware Anchor Points} 
Our primary optimization strategy for 3D Gaussians involves gradually removing points irrelevant to facial motion and duplicating those that contribute significantly (termed as anchor points), as shown in Figure~\ref{fig_simpliciation} (a). For static scene or object reconstruction, the gradient accumulation of each 3D Gaussians position in the optimizer serves as the basis for duplication and pruning. However, this point cloud optimization method is not suitable for our task, as the regions around the eyes and teeth is dynamic. \par 

Meanwhile, we initialize a learnable binary auxiliary Gaussian attribute from the UV vertices, named as \textbf{Idx} $\in \{0,1\}^{N}$ ($N$ is the number of points), as shown in Figure~\ref{fig_pipeline}. The accumulation of density gradients does not only come from the optimized gradients but also from positional differences relative to the normalized canonical point cloud. In practice, we apply the canonical point cloud ($\bar{\textbf{P}}_0$) of the face in the first frame as the normalized reference. Then for each Gaussian point, we calculate the gradient as the following: 
\begin{equation}
\label{eq6}
\mathit{\mathbf{Idx}} = \left\{
\begin{aligned}
1 & , & <\alpha_{grad}, \nabla \mathbf{P}>_{\text{max}} \geq \epsilon, \\
0 & , & <\alpha_{grad}, \nabla \mathbf{P}>_{\text{max}} < \epsilon.
\end{aligned}
\right.
\end{equation}
Here, $\epsilon$ represents the threshold for the maximum between optimizer and motion gradient, the $\nabla \mathbf{P}$ is the accumulated gradient of $\mathbf{P}$ relative to $\bar{\textbf{P}}_0$ (first-order difference). Points with a value of $1$ in \textbf{Idx} will be cloned, while those with a value of $0$ will be pruned. The $\alpha_{grad}$ is the gradient of density. \par

This approach retains points that contribute to facial motion while eliminating those do not, as in Figure~\ref{fig_simpliciation} (b). These motion anchors effectively control the proliferation of non-contributory points and adapt swiftly to facial expression changes without requiring additional neural networks (\textit{e.g.}, MLPs or CNNs) for fitting facial motion via expression. However, due to duplication, the excessive density around the anchor points leads to additional computational overhead as the anchor points and their duplicates proliferate exponentially.

\subsection{Gaussian Points Simplification} 
To regulate the over-proliferation of Gaussian anchor points, the basic strategy periodically reduces opacity values, removing points that consistently remain at low opacity levels. However, this opacity-based control method is effective in static scene reconstruction, but it conflicts with our method. Generally, points that are fully transparent (with opacity $0$) typically do not contribute to motion, such as those located on the forehead around the head. Then, we propose two methods for optimizing the number of these points, removing excessively small points and aligning remaining points closer to geometric surface. Points positioned closer to surface better represent details, while those farther away tend to introduce noise.\par 

We apply the learnable binary index $\mathbf{Idx}$ as the mask on the opacity property $o$ and scale property $s$ (updated to $o',s'$), as: 
\begin{equation}
\begin{aligned}
[s',o'] = \mathbf{Idx} \cdot [s,o],
\end{aligned}
\end{equation}
which means that those Gaussian points with motion gradient contributions ($\mathbf{Idx}$ value with $1$) are additionally applied on the scale and opacity. Then, we adopt the straight-through estimator~\cite{bengio2013estimating,lee2024compact,van2017neural} for gradient calculation with binary parameters, as the following: 
\begin{equation}
\begin{aligned}
\mathbf{Idx}' = \text{sigm}(\mathbf{Idx}) + \text{sg}[\mathbb{U}[\text{sigm}({\mathbf{Idx}}) > \epsilon'] - \text{sigm}(\mathbf{Idx})],
\end{aligned}
\end{equation}
where the $\text{sigm}$, $\text{sg}$, $\mathbb{U}$ are the sigmoid function, stop gradients and indicator function (mapping with $0$ and $1$). The $\epsilon'$ are the pre-defined threshold, which is set to $0.01$. During forward-propagation, the mask applied on $o$ and $s$ via the value of $\mathbf{Idx}$. In back-propagation, gradients are obtained from the derivative of $\text{sigm}(\mathbf{Idx})$, which solves the non-differentiability of binarization. It is shown in Figure~\ref{fig_simpliciation} (c), the redundant points are eliminated. \par 

Meanwhile, we propose a point-to-surface measurement to reduce the proliferation of points located far from the face, as these points contribute little yet significantly increase computational cost. Specifically, the projected distance of each point to the initialization surface is calculated, and the positions of discrete points and their corresponding surface points are indexed by the the anchor points and point clusters bound by the anchor points in:  
\begin{equation}
\label{8}
\begin{aligned}
d = \sum_{A \in \mathbf{Idx}} \mathbf{Idx} \cdot \sum_{\mathbf{P}\in A} ||(\mathbf{P}-\mathbf{P}_0^A)\cdot n_0^A||,
\end{aligned}
\end{equation}
where $n_0^A$ is unit normal of anchor points $\mathbf{P}_0^A$, and $\mathbf{P}$ is the bound points with anchor $\mathbf{P}_0^A$, then we find the anchor points through $\mathbf{Idx}$. The regularization on $d$ will make the bound points around the anchor point closer to the corresponding position on the surface, and minimizing the offset. 

\section{Experiments}

\subsection{Implementation Details}

\subsubsection{Datasets.} We perform experiments with $6$ subjects monocular videos from the public datasets GaussianBlendshape~\cite{ma20243d}, NeRFBlendshape~\cite{Gao2022nerfblendshape}, StyleAvatar~\cite{wang2023styleavatar}, and InstantAvatar~\cite{zielonka2023instant} in the offline training setup. Moreover, we apply self-captured video via webcam within on-the-fly training. The online/offline experiments are presented with the resolution of $512 \times 512$. In offline setup, we take an average length of $1000$-$5000$ frames for training (about 80\%) while the test dataset includes frames with novel expressions and poses (about 20\%), which is aligned with baseline methods. The sequential/random inputs are applied for online/offline set-up, respectively. For each frame, the RobustVideoMatting~\cite{lin2022robust} is used to remove background. 

\subsubsection{Hyperparameters.} The 3D Gaussian points are initialized with learning rates for Gaussian properties $\{\alpha, s, r, o, \mu, \mathit{Idx}\}$ set to $\{1e^{-5}, 5e^{-3}, 1e^{-3}, 5e^{-2}, 2.5e^{-3}, 1e^{-4}\}$. During pre-training, the learning rates for auxiliary properties $\{\mathit{Tex}, \mathit{SH}\}$ are $\{2e^{-4}, 1e^{-4}\}$, while the others remain the same. The threshold of $\alpha_{grad}$ and $\nabla \mathbf{P}$ is set to $0.01$. The $\alpha_{grad}$ is computed by the accumulated gradient of $\alpha$ (the self.xyz\_gradient\_accum in 3DGS). For the online optimization, we construct each mini-batch with multiple images, comprising one newly captured frame and several previously captured frames, to mitigate forgetting of historical information. 

\subsubsection{Losses.} We apply $L_1$ and SSIM in training. Meanwhile, the distance between the points and surface in Eq.~\ref{8} as regularization. For warm-up phrase, we introduce dark channel loss to separate the learnable $\mathit{Tex}$ and $\mathit{SH}$ coefficients. During training, the weights for the $L_1$ and SSIM losses are set to 1 and 0.1, respectively, with a regularization term weight of $0.01$. Additionally, a dark channel loss with the weight of 10 is incorporated for the warm-up phrase.

\subsubsection{Pipeline Details.} We set the head geometry corresponding to the initial frame as canonical points clouds, and the position difference of the subsequent point clouds relative to the canonical point cloud is used as the motion gradient. The motion gradients are also cloned with the anchor points and deleted with non-contributing points. We achieve that by introducing learnable parameters $\mathbf{Idx}$. The points with $\mathbf{Idx}$ value corresponds to $0$ are deleted, and the points with value as $1$ are cloned. We allocate approximately 30 seconds for the warm-up phase within on-the-fly pipeline (a longer warm-up duration is not recommended, as it may cause collapse). We execute the cloning/pruning every 1500 optimization iterations. 



%

\begin{figure}[t]
  \centering
  \includegraphics[width=.99\linewidth]{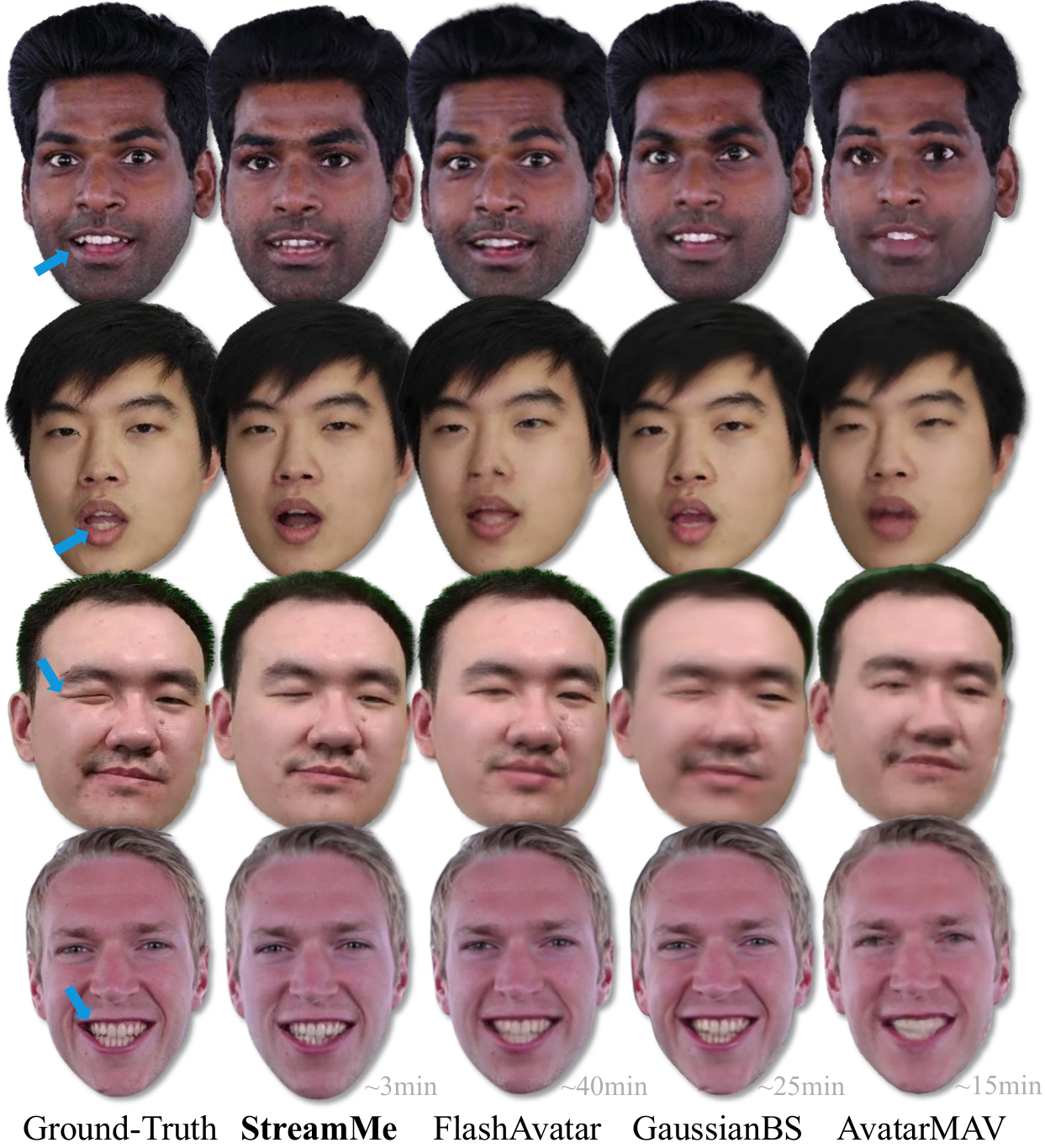}
  \vspace{-0.2cm}
  \caption{The perceptual evaluation of our method and baselines for self-reenactment. Baseline methods are evaluated with their default settings, and time consumption is recorded under the official provided iteration. Blue arrows on the ground-truth highlight regions of interest for closer inspection. Please zoom in for detailed comparison. Natural face$\copyright$\textit{Lizhen Wang et al.} (CC BY), and $\copyright$\textit{Wojciech Zielonka et al.} (CC BY).}
  \vspace{-0.4cm}
  \label{fig:compare}
\end{figure}

\subsection{Baseline}
Given the challenges associated with high-efficient avatar reconstruction, especially completing reconstruction within minutes while achieving re-animation, few methods exactly handle this task. Therefore, we select several related works for comparison, as,
\begin{itemize}
\setlength{\itemsep}{5pt}
\setlength{\parsep}{0pt}
\setlength{\parskip}{0pt}
\item \textbf{AvatarMAV}~\cite{xu2023avatarmav}: This method applies a NeRF-based implicit neural blend representation. By training a lightweight-MLP, it integrates multiple learnable implicit neural shapes for appearance. It claims to accomplish head reconstruction in 5 minutes ($256 \times 256$ resolution). Here, the resolution is reset to $512 \times 512$ to align with baselines.
\item \textbf{FlashAvatar}~\cite{xiang2023flashavatar}: The FlashAvatar builds on 3DGS, utilizing UV sampling for Gaussian initialization and an offset network with MLPs to dynamically model variations in facial expressions. It reports inference speed but omits the time required for reconstruction.
\item \textbf{GaussianBlendshape}~\cite{ma20243d}: It is the state-of-the-art in monocular reconstruction based on 3DGS. This approach models facial motion through the mixture of explicitly learnable blendshapes aligned with the pre-tracked FLAME expression coefficients. 
\end{itemize}
The FlashAvatar and GaussianBlendshape are from the preprocessed FLAME via MICA~\cite{MICA:ECCV2022}, which is time-intensive and unsuitable for on-the-fly reconstitution. We also acknowledge other related methods, such as NeRFBlendshape~\cite{Gao2022nerfblendshape}, HeadGaS~\cite{dhamo2024headgas} and MonoGaussianAvatar~\cite{chen2024monogaussianavatar}~\textit{etc}. However, we exclude these methods from comparison due to their on-the-fly reconstruction setups and comparable performance (\textit{e.g.}, INSTA~\cite{zielonka2023instant} and AvatarMAV employ fast training by sampling rays from NeRF~\cite{mildenhall2020nerf}, yet exhibit slow inference when releasing all rays).


\subsection{Numerical Results and Comparisons}

We take two criteria for numerical evaluations, one is from quantitative measurement, the other is from human assessment. 

\subsubsection{Quantitative Metrics} It is based on three aspects. (1) Image Quality: We use the PSNR, LPIPS~\cite{zhang2018perceptual} and MSE for the evaluation of self-reenactment image quality. (2) Inference Frame Rate: The inference frame rate (FPS) the measurement of the number of frames generated within per second without introducing the head tracking, which is not the speed of the pipeline. It is measured on a single NVIDIA RTX4090 GPU. (3) Memory Storage: The memory storage (\textit{Mem.}) is the storage capacity occupied by the models, the compact models offer advantages in both computational efficiency and storage requirements. We use the Megabyte (\textit{MB}) as units. The quantitative experiments are performed on the self-reenactment. 

\subsubsection{User Study} We sample $6$ distinct identities, each represented by $20$ video clips ($5$ for self-reenactment and $15$ for cross-reenactment), and invite $30$ participants for the human evaluation. The Mean Opinion Scores (MOS) rating protocol is employed, with participants asked to assess the generated videos across four criteria: (1) \textbf{MS} (Motion Synchronization): To what extent do you agree that the head motion in the animated videos is synchronized with the driving source? and (2) \textbf{VQ} (Video Quality): To what extent do you agree that the overall video quality is high, considering factors such as frame quality, temporal consistency, and so forth? A 5-point Likert scale is used for each criterion, with scores ranging from $1$ to $5$, where $1$ represents "strongly disagree" and $5$ represents "strongly agree" (higher scores indicate better performance). 

\begin{figure}[t]
  \centering
  \vspace{-0.4cm}
  \includegraphics[width=.99\linewidth]{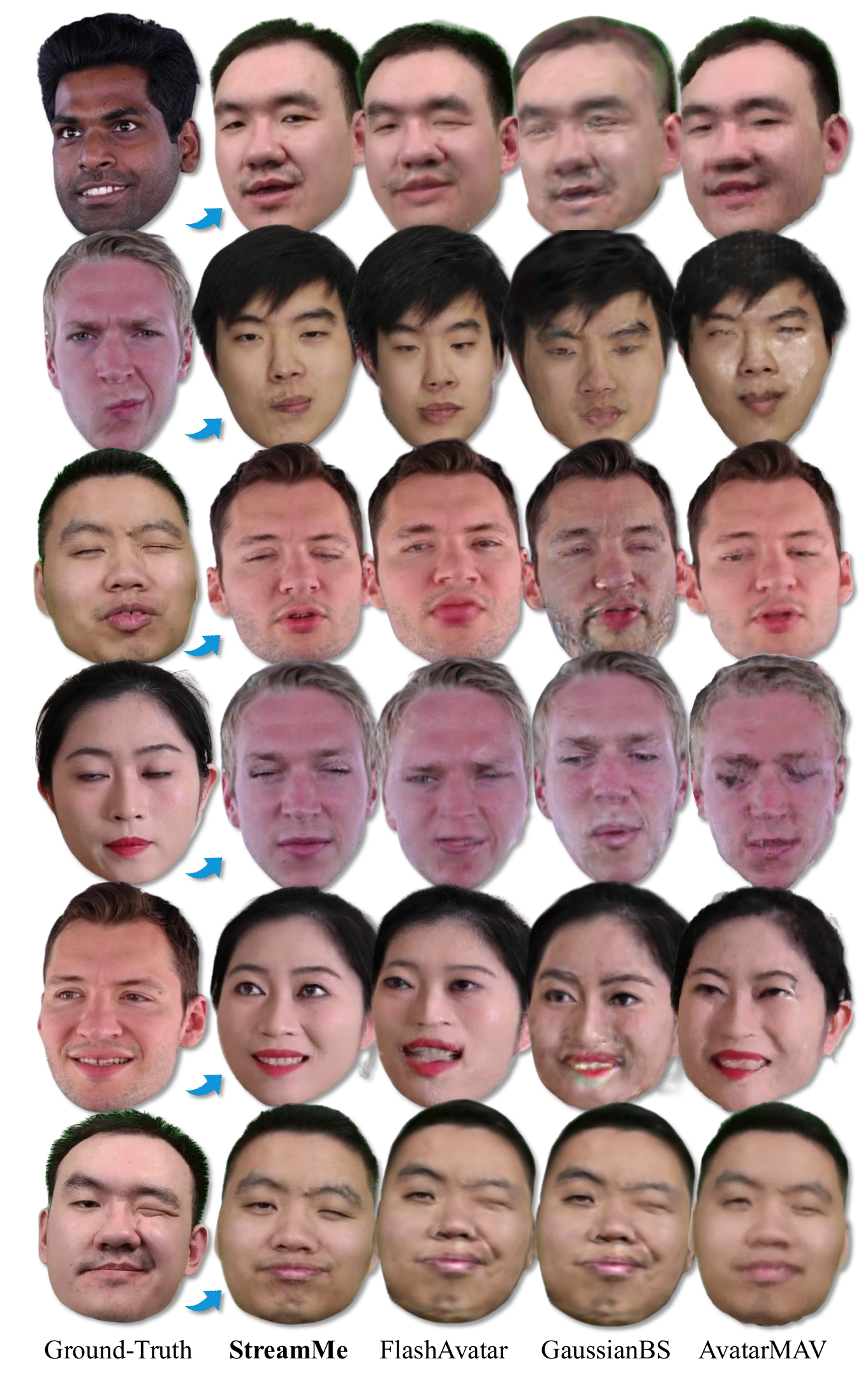}
  \vspace{-0.3cm}
  \caption{The perceptual evaluation of our method and baselines for cross-reenactment. Our method demonstrates superior preservation of high-frequency facial details, including features such as teeth and hair. Please zoom in for a closer view. Natural face$\copyright$\textit{Lizhen Wang et al.} (CC BY), $\copyright$\textit{Xuan Gao et al.} (CC BY) and $\copyright$\textit{Wojciech Zielonka et al.} (CC BY).}
  \label{fig:cross}
\end{figure}

\begin{table}[t]
\footnotesize
\begin{center}
\caption{(1) Left part: Quantitative evaluation results compared with baseline methods. The best scores are highlighted in bold, with the second-best underlined. Symbols $\downarrow$ and $\uparrow$ denote whether lower or higher values indicate superior performance, respectively. All experiments were conducted on a single NVIDIA RTX4090 machine. (2) Right part: A 5-point Likert scale is used for the user study, where scores closer to 5 signify better performance.}
\vspace{-0.3cm}
\setlength{\tabcolsep}{1.2mm}
{
\begin{tabular}{cccccc|cc}
\hlinew{.8pt}
\multirow{3}{*}{Methods} &PSNR$\uparrow$ &MSE$\downarrow$ &LPIPS$\downarrow$  &FPS$\uparrow$ &\textit{Mem.}$\downarrow$ &\textbf{MS}$\uparrow$ &\textbf{VQ}$\uparrow$ \\
\cline{7-8}
&dB &$\rightarrow 0$ &$\rightarrow 0$ &  & MB &\multicolumn{2}{c}{$\rightarrow 5$}\\
\cline{2-8}
&\multicolumn{5}{c|}{Quantitative Results}&\multicolumn{2}{c}{User Study}\\
\hline
AvatarMAV & 24.1 & 0.047 & 0.137  & 2.58  & {14.1} & 3.1 & {2.3}  \\
FlashAvatar & \underline{27.8} & 0.021 & \underline{0.109} & \underline{94.5} & \underline{12.6}  &   \underline{3.8} & 3.1    \\
GaussianBlendshape &26.4& \underline{0.017} & 0.112 & 22.9 &  872 &  3.7 & \underline{3.6}  \\
\rowcolor{mygray} \textbf{StreamME} &\textbf{29.7} & \textbf{0.012} &\textbf{0.095} &\textbf{139}  &\textbf{2.52} & \textbf{3.9} & \textbf{4.1} \\
\hlinew{.8pt}
\end{tabular}}
\vspace{-0.5cm}
\label{table_1}
\end{center}
\end{table}

\subsection{Quality Comparison with Baseline Methods}

The quality comparison results of self-reenactment are shown in Table~\ref{table_1} and Figure~\ref{fig:compare} respectively. From Table~\ref{table_1}, our method achieves the best results in user study and quantitative evaluations. The FlashAvatar is a powerful baseline, but still requires a lot of time to train the learnable MLP layers. The AvatarMAV is fast but exhibits limited detail preservation at $512\times 512$ resolution. 

The perceptual comparisons for cross-reenactment are illustrated in Figure~\ref{fig:cross}, where our method consistently delivers superior results. The AvatarMAV and FlashAvatar exhibit noticeable artifacts on out-of-distribution expressions, as their learnable MLP layers struggle to adapt to expression changes within few iterations (for efficiency). GaussianBlendshape, on the other hand, underperforms due to the lack of conditions aligned with blendshapes during training. Our approach circumvents the need for MLPs and blendshapes as training conditions by directly binding appearance representations to the point cloud, enabling efficient training while preserving high quality across diverse facial expressions.


%
\begin{table}[t]
\footnotesize
\begin{center}
\vspace{-0.3cm}
\caption{The quantitative results of image quality and number of iterations (\textit{Iters.}) at each time point. The \textit{Iters.} represents the training iteration speed in the same unit time, a higher value indicates better training efficiency. Our method achieves comparable quality in just two minutes, whereas other methods require 30 minutes to reach the same standard.}
\vspace{-0.4cm}
\setlength{\tabcolsep}{1.0mm}
{
\begin{tabular}{ccc|cc|cc|cc}
\hlinew{.8pt}
\multirow{2}{*}{Methods} &PSNR &\textit{Iters.} &PSNR  &\textit{Iters.} &PSNR &\textit{Iters.} &PSNR &\textit{Iters.} \\
\cline{7-9}
\cline{2-9}
&\multicolumn{2}{c|}{Time = 1s}&\multicolumn{2}{c|}{Time = 10s}&\multicolumn{2}{c|}{Time = 2min}&\multicolumn{2}{c}{Time = 30in}\\
\hline
AvatarMAV & 3.12 & 7.6$e^1$ & 21.7  & 7.9$e^2$  & 24.1 & 9.6$e^3$ & 24.4 & 1.4$e^5$  \\
FlashAvatar & 4.90 & 9.7$e^1$ & 12.5 & 9.4$e^2$ & 16.9  & 1.2$e^4$ & 26.8  & 1.7$e^5$  \\
GaussianBlendshape &7.29& 2.9$e^1$ & 15.6 & 3.7$e^2$ &  25.2 &  4.8$e^3$ & 25.8 & 6.1$e^4$ \\
\rowcolor{mygray} \textbf{StreamME} &\textbf{10.8} & \textbf{1.4$e^2$} &\textbf{23.1} & \textbf{1.5$e^3$}  &\textbf{27.2} & \textbf{1.6$e^4$} & \textbf{29.8} & \textbf{3.4$e^5$} \\
\hlinew{.8pt}
\end{tabular}}
\label{table_2}
\end{center}
\end{table}

\subsection{Efficiency Comparison with Baseline Methods}

\begin{figure}[t]
  \centering
  \includegraphics[width=1.\linewidth]{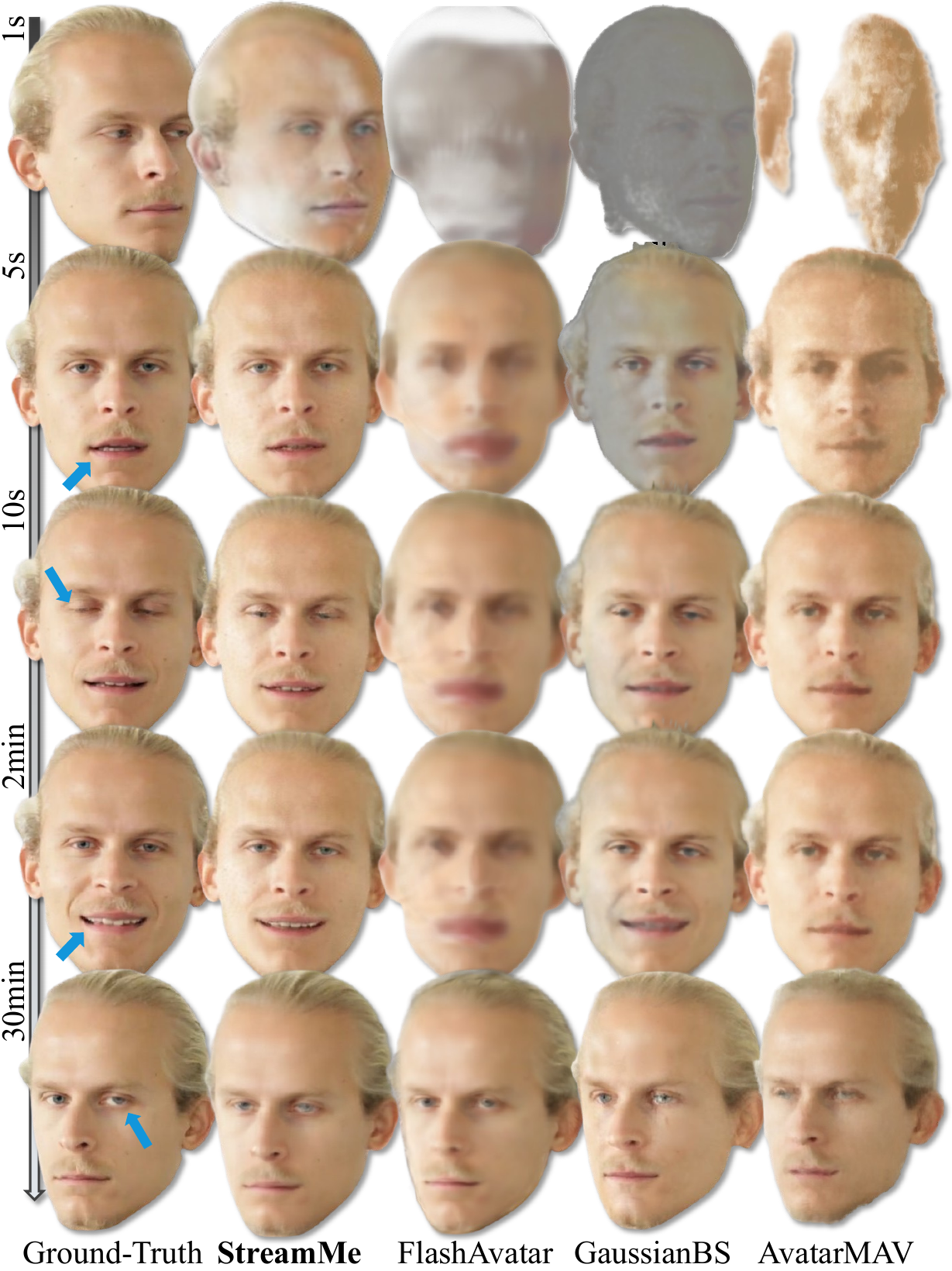}
  \caption{The visualization of efficiency comparison with baseline methods, with results recorded from 1 second until the point at which all methods achieve convergence (about 30 minutes). Please zoom in to compare the fine details of teeth, eyes, hair, and the subtle differences in facial expressions. It is noteworthy that our method is able to achieve the representation of expressions and details with few iterations and time consumption. Natural face $\copyright$\textit{Wojciech Zielonka et al.} (CC BY).}\label{fig:time-compare}
\end{figure}

In addition to quality comparisons, we further validate the efficiency improvements achieved by our method. 
Specifically, we evaluate the models at various training stages (iterations and time) on testset. The model capacity at each training slot will be examined for comparison. The results are shown in Figure~\ref{fig:time-compare} and Table~\ref{table_2}, it can be found that compared with the baseline method, our approach achieves convergence within 2 minutes, and simulates dynamic expression from the outset. As shown in Figure~\ref{fig:time-compare}, our method synthesizes detailed tooth within just 10 seconds and simulates dynamic expressions without any warm-up phase. The efficiency is due to it is geometric foundation, avoiding reliance on learnable MLPs. Additionally, as shown in the Table~\ref{table_2}, different from the nearly constant training speed of baseline methods, our method improves training efficiency over time due to the progressively sparse point clouds, which reduce computational redundancy.

\subsection{Ablation Study}

\begin{figure}[htp]
  \centering
  \includegraphics[width=1.\linewidth]{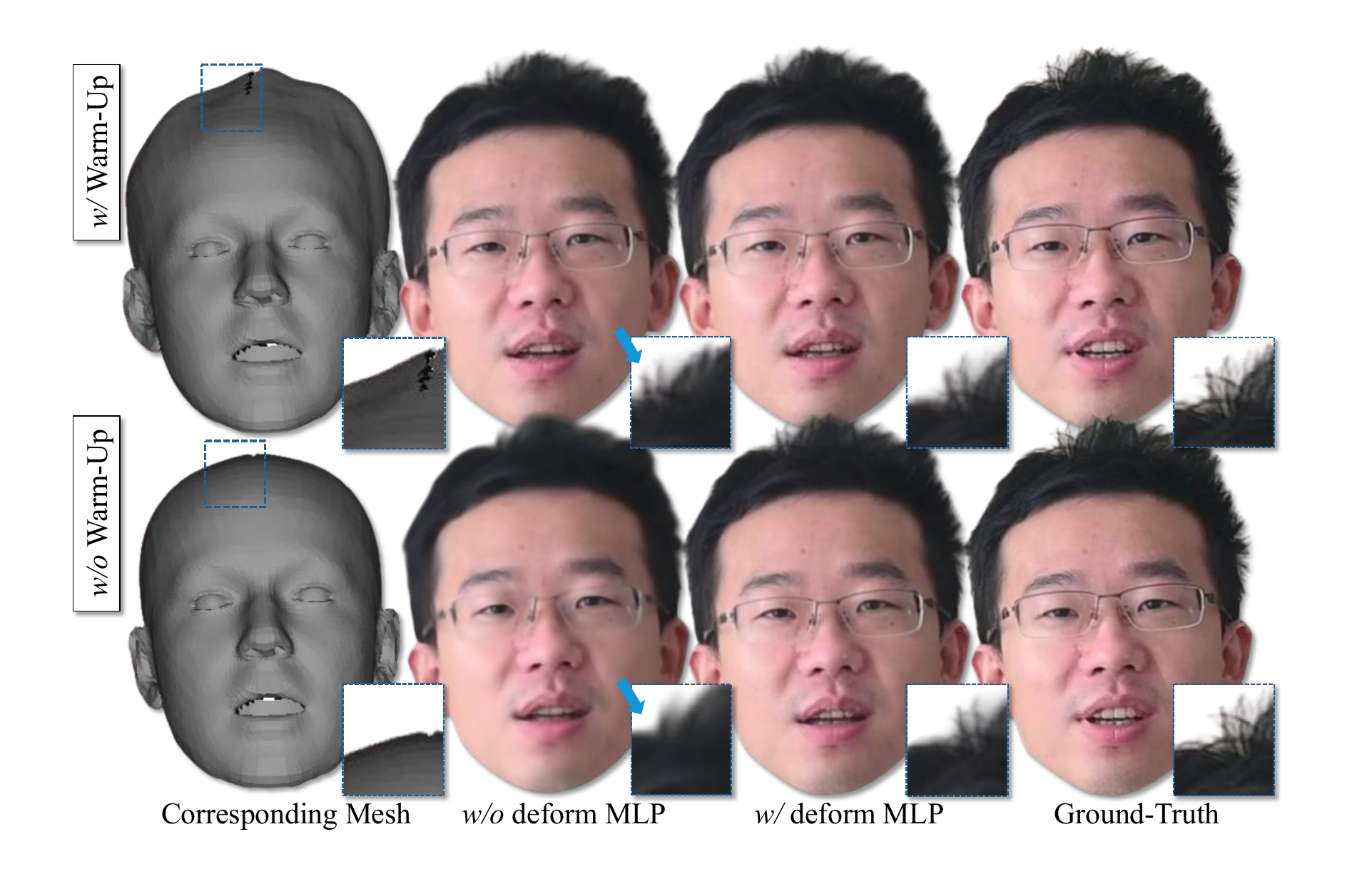}
  \vspace{-0.2cm}
  \caption{The visualize of ablation study on Gaussian properties warm-up phrase. Here, the explicit geometric deformation serves the same function as the deform MLP during the 3DGS training phase. For results without the deform MLP, those initialized with accurate geometric deformation from the warm-up phase show improved quality over those initialized from the template head, particularly in outer facial regions like the hair via red arrow, as seen in the second column (left to right). However, introducing the MLP in 3DGS training compensates effectively for this difference, as illustrated in the third column (left to right), since both approaches focus on adjusting point positions. Natural face$\copyright$\textit{Luchuan Song et al.} (CC BY).}
  \vspace{-0.1cm}
  \label{fig:ablation 1}
\end{figure}

\begin{figure}[htp]
  \centering
  \includegraphics[width=.99\linewidth]{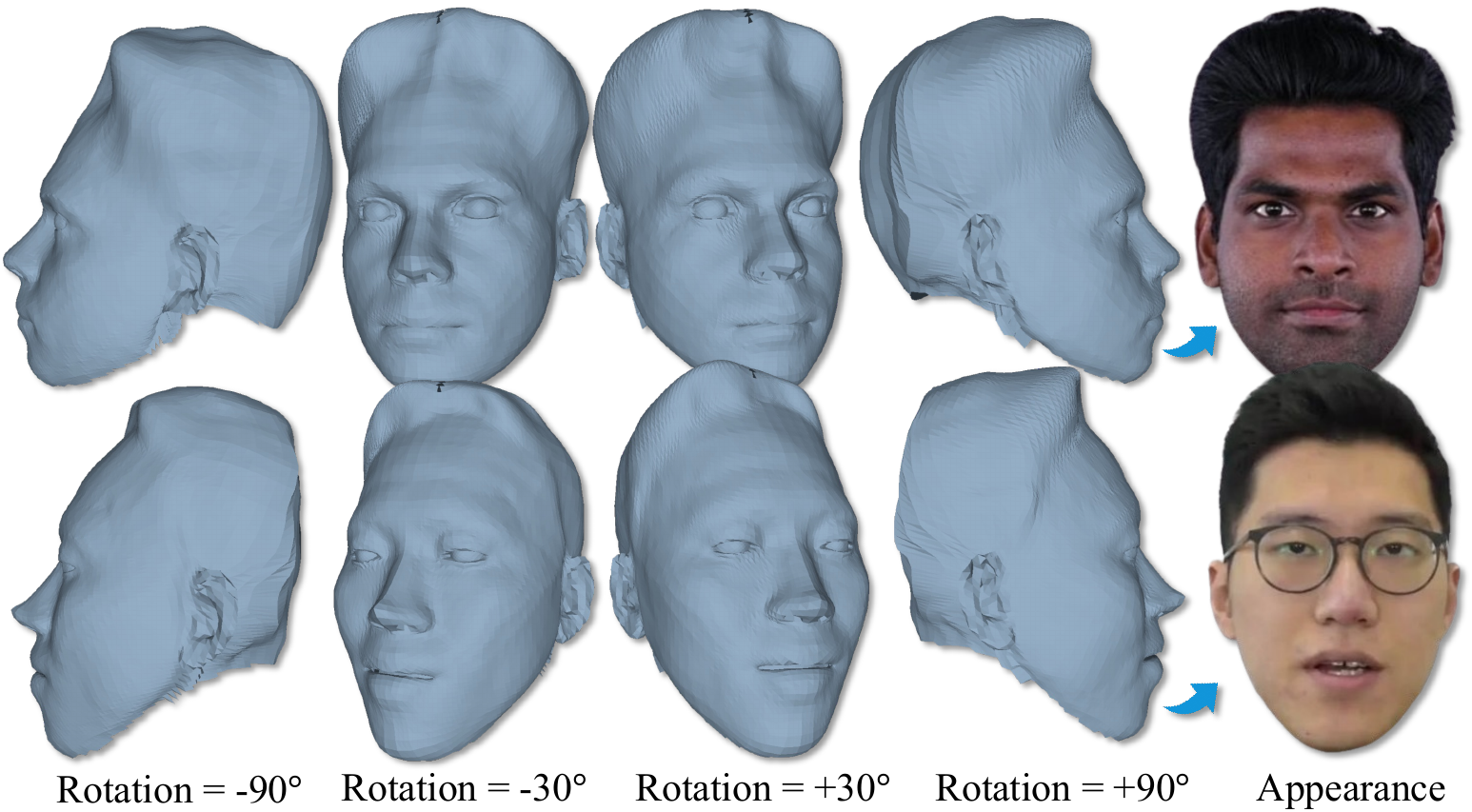}
  \vspace{-0.1cm}
  \caption{The warm-up phase helps geometry adjustment for 3D Gaussian initialization. We showcase its head geometry simulation from multiple viewpoints, highlighting the consistent of hair geometry. The vertex offset obtained could subsequently be applied to improve the head tracking. Natural face$\copyright$\textit{Wojciech Zielonka et al.} (CC BY).}
  \vspace{-0.2cm}
  \label{fig:ablation 2}
\end{figure}

\begin{figure}[htp]
  \centering
  \includegraphics[width=.95\linewidth]{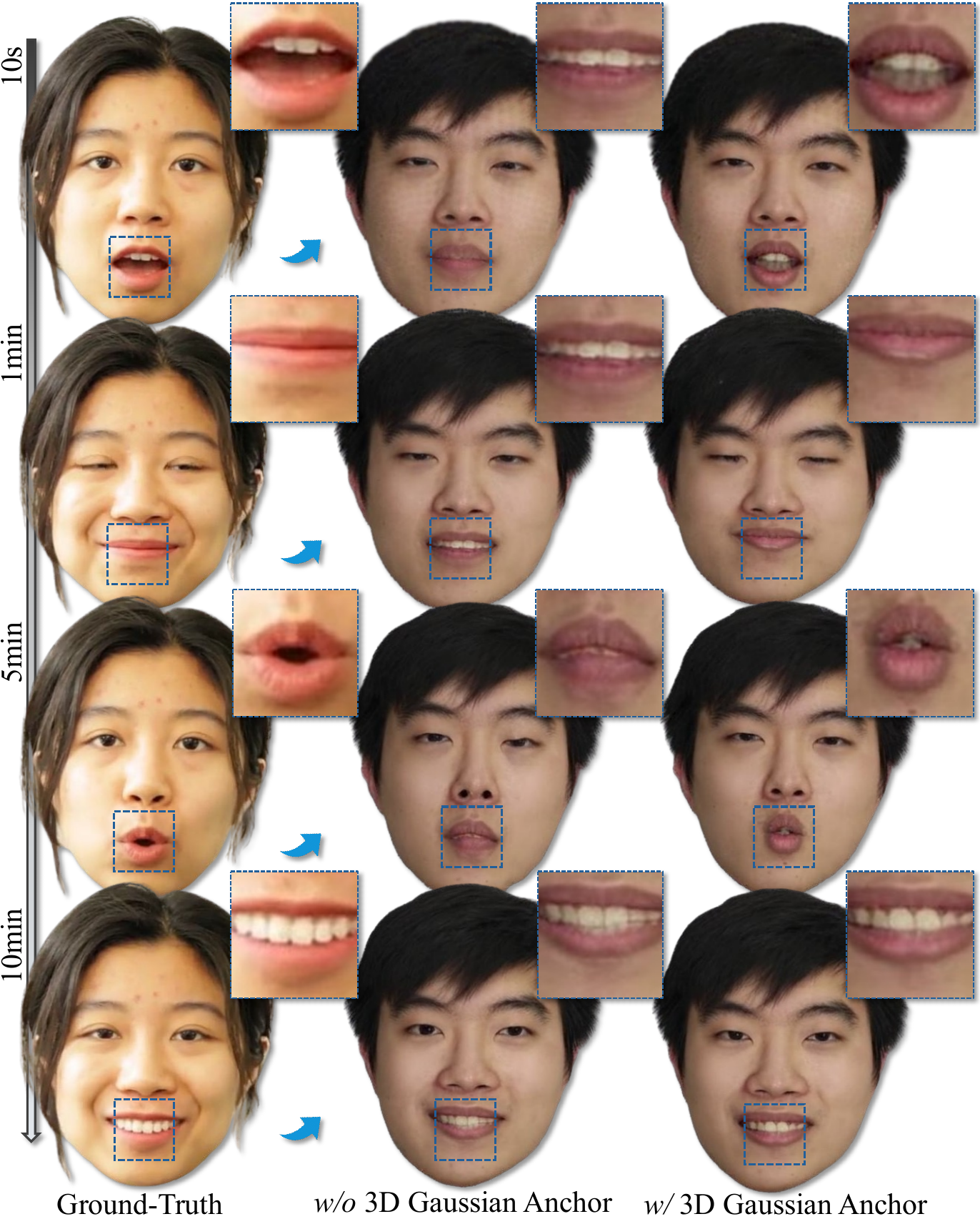}
  \vspace{-0.4cm}
  \caption{The visualize of image quality via 3D Gaussian anchor. We demonstrate this through cross-reenactment perceptual evaluation. With the 3D Gaussian anchor, the model instantly adapts to dynamic expressions. We highlight the shape of the mouth for comparison. Natural face$\copyright$\textit{Yufeng Zheng et al.} (CC BY) and $\copyright$\textit{Lizhen Wang et al.} (CC BY).}
  \vspace{-0.3cm}
  \label{fig:ablation 3}
\end{figure}

\begin{figure}[htp]
  \centering
  %
  \includegraphics[width=.92\linewidth]{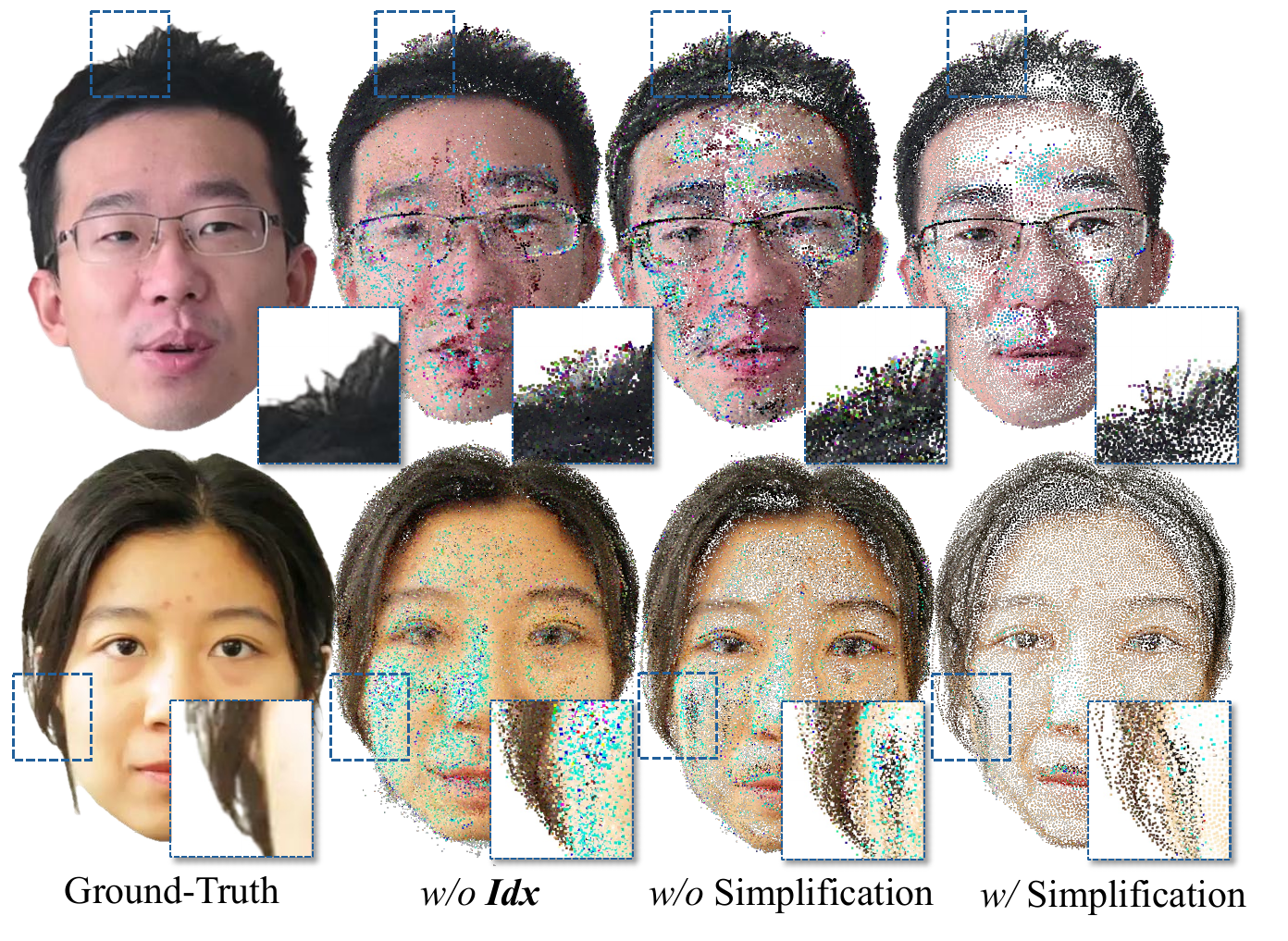}
  \vspace{-0.4cm}
  \caption{The visualize of points cloud visualization via 3D Gaussian anchor and simplification. By utilizing the gradient with the anchor points, the point cloud is filtered to eliminate points with low contributions to motion. This simplification further optimizes the number of point clouds. As shown by the detailed representation of hair, the optimized point clouds preserve high-frequency features. Natural face$\copyright$\textit{Yufeng Zheng et al.} (CC BY).}
  \vspace{-0.3cm}
  \label{fig:ablation 4}
\end{figure}

In this section, we present ablation studies on Gaussian properties warm-up, Gaussian motion anchors, and Gaussian simplification to validate the importance of these modules.


\subsubsection{Ablation study on Gaussian properties warm-up} We present the improvement provided by the 3D Gaussian properties warm-up. As illustrated in Figure~\ref{fig_pipeline}, we label it as optional since it functions equivalently to the point cloud deformation MLP. Specifically, it pre-simulates the point cloud offset of the 3D head template, providing prior positional information for the 3D Gaussians. To conduct the ablation study, we introduce the position offset MLP (delta MLP) with (\textit{w/}) and without (\textit{w/o}) warm-up intervention, as shown in Figure~\ref{fig:ablation 1}. In the results without the delta MLP, warm-up contributes to clearer details in out-of-face areas, such as hair. Incorporating the MLP mitigates this issue, improving the detail clarity, while slightly reduce execution efficiency.

At the same time, the properties warm-up will provide head geometry, which only takes about 10 seconds and several appearances from different perspectives for fitting, as shown in Figure~\ref{fig:ablation 2}, which is much faster than head fitting algorithms such as VHAP~\cite{qian2024versatile} and MonoNPHM~\cite{giebenhain2024mononphm}. Although its geometric accuracy could be further enhanced, it strikes a balance between geometry and 3D Gaussian feature representation.

\begin{figure}[t]
  \centering
  \includegraphics[width=1.\linewidth]{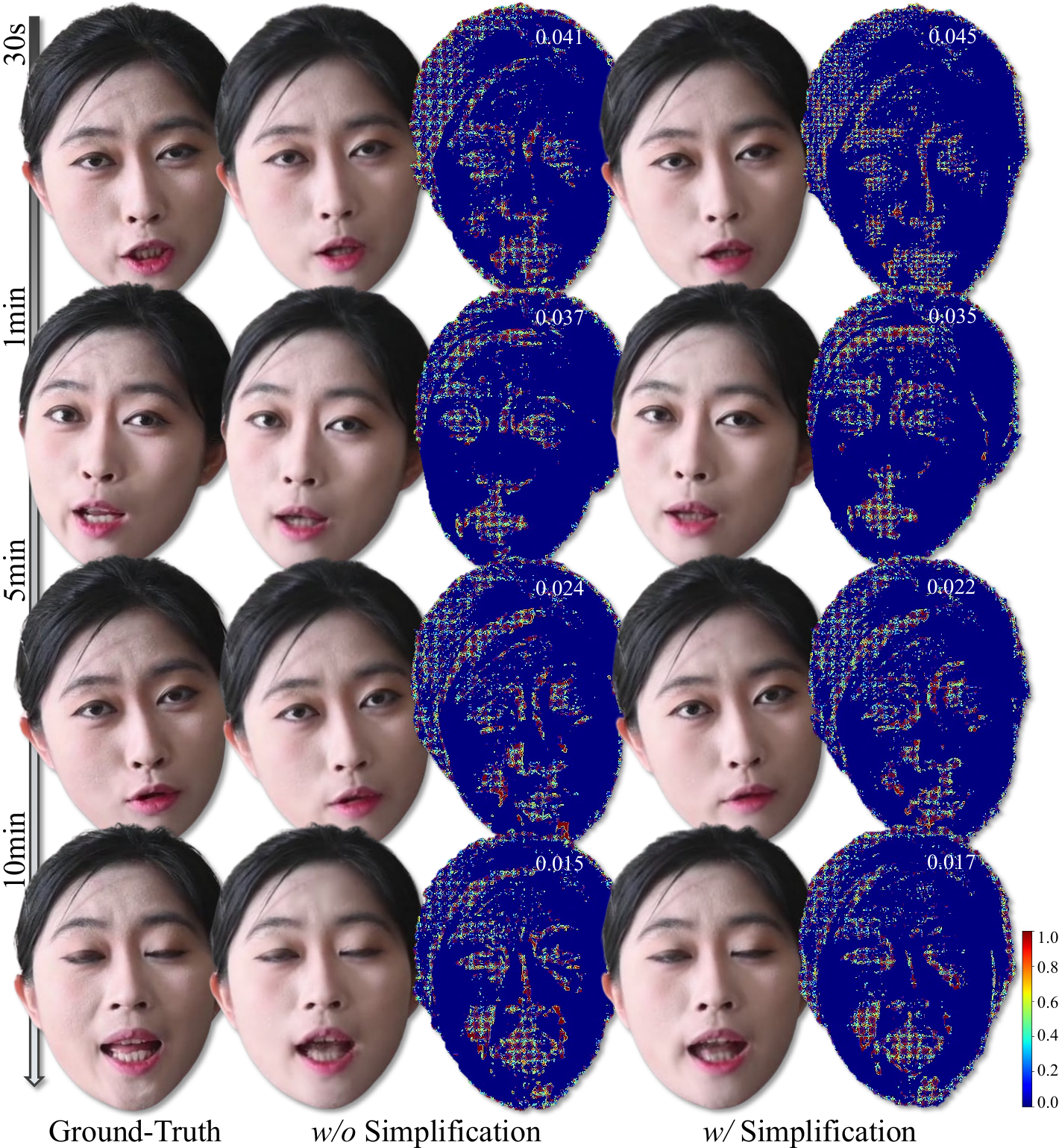}
  \vspace{-0.5cm}
  \caption{The visualize of the 3D Gaussian simplification ablation study. We present the rendered appearance and MSE pixel error maps corresponding to with (\textit{w/}) and without(\textit{w/o}) simplification at different time slots. The mean error values are in the upper right corner of each error map. As the first pruning does not occur at 0 sec, we begin recording from the 30th second. The introduction of simplification will not result in any reduction in qualitative and quantitative results, but lead to a significant improvement in efficiency. Natural face$\copyright$\textit{Luchuan Song et al.} (CC BY).}
  \label{fig:ablation 5}
\end{figure}

\begin{figure}[h]
  \centering
  \vspace{-0.1cm}
  \includegraphics[width=1.\linewidth]{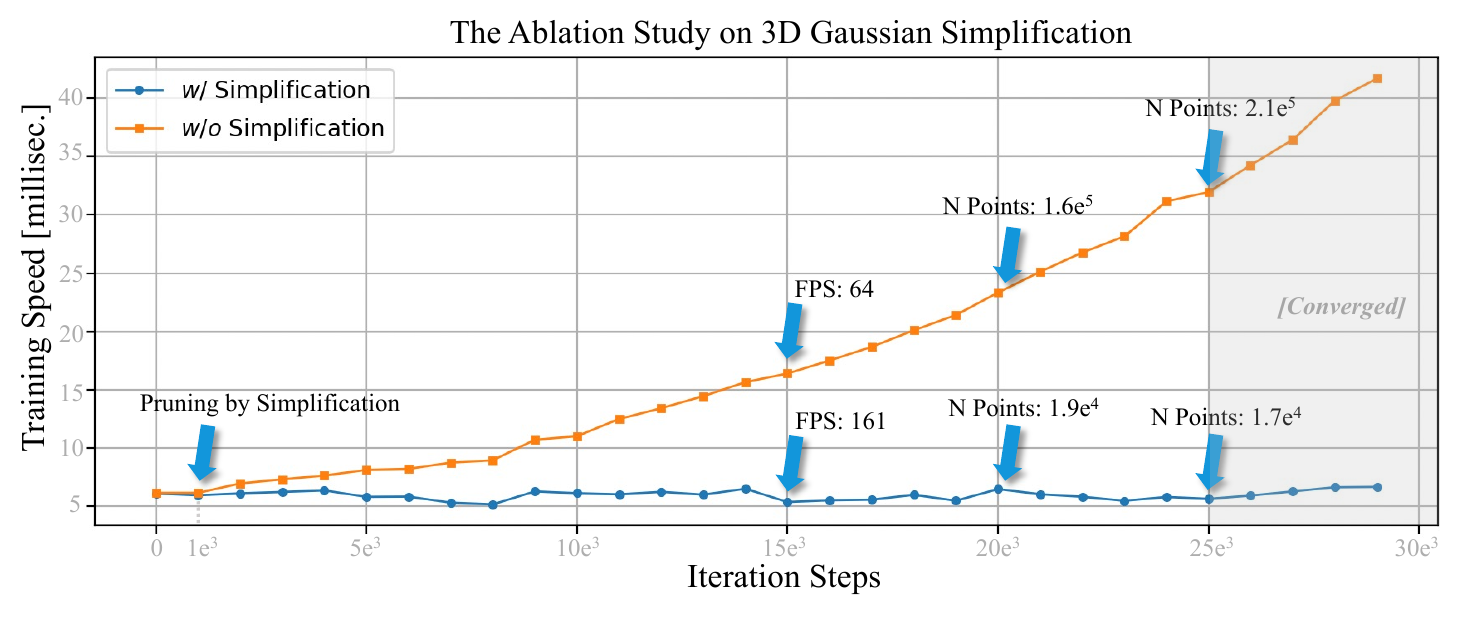}
  \vspace{-0.5cm}
  \caption{Comparisons in the ablation study on \textit{w/} and \textit{w/o} 3D Gaussian simplification. For \textit{w/o} simplification, as the number of iterations grows, the time per iteration also increases due to the overhead from redundant points. In contrast, simplified pruning (\textit{w/} simplification) enables the system to maintain high efficiency throughout. The FPS represents the training iterations per second. The corresponding number of 3D Gaussian points are annotated for reference. And the dark gray region indicates that \textit{w/} simplification has converged.}
  \label{fig:ablation 6}
\end{figure}

\subsubsection{Ablation study on motion-aware anchor} 
The motion-aware anchor selection leverages the motion gradient prior, progressively duplicating points associated with motion. This approach enables dynamic facial expressions to be directly mapped onto geometric structures, bypassing the need for extensive iterative training steps for adaptation. As shown in Figure~\ref{fig:ablation 3}, the motion-aware anchor efficiently adapts to dynamic facial expressions at the beginning of training, whereas without it, a warp-up period of approximately 10 minutes would be required. Additionally, it optimizes the number of point clouds by incorporating motion gradients, and eliminates the points unrelated to facial movement, as illustrated in Figure~\ref{fig:ablation 4}. 



\subsubsection{Ablation study on 3D Gaussian simplification} 
The simplification of 3D Gaussian points is designed to reduce computational load without sacrificing rendering quality, as shown in Figure~\ref{fig:ablation 4} (\textit{w/} Simplification). Although denser point clouds are generally associated with higher detail, in our case, redundant points contribute minimally to 3DGS. Figure~\ref{fig:ablation 5} presents the comparison of results with (\textit{w/}) and without (\textit{w/o}) simplification. As shown, point cloud simplification does not result in a decrease in quality, as confirmed by the error maps and mean error values. Specially, after more than 10 minutes of training (about $1e^5$ iterations), \textit{w/} simplification achieves approximately $8-10$ times the reduction of points number compared to \textit{w/o} simplification (the numbers are from $200k$ to $24k$), which has a significant improvement in efficiency. Additionally, the quantitative comparison results are shown in the Figure~\ref{fig:ablation 6}, the method with simplification (\textit{w/} simplification) maintains high efficiency throughout the training phase. Moreover, the average number of point clouds for training subjects after convergence is 9,807, compared to 13,453 for FlashAvatar and 62,530 for GaussianBlendshape, which demonstrates the cloning and pruning effectively reduce the number of points.

\section{Applications}
\label{Applications}

We extend our method to support a series of downstream applications in addition to cross-reenactment (facial relighting and toonification). We introduce these applications with streamME as the foundation and provide baselines methods for comparison.

\subsection{Facial Relighting Application}

\begin{figure}[t]
  \centering
  \includegraphics[width=1.\linewidth]{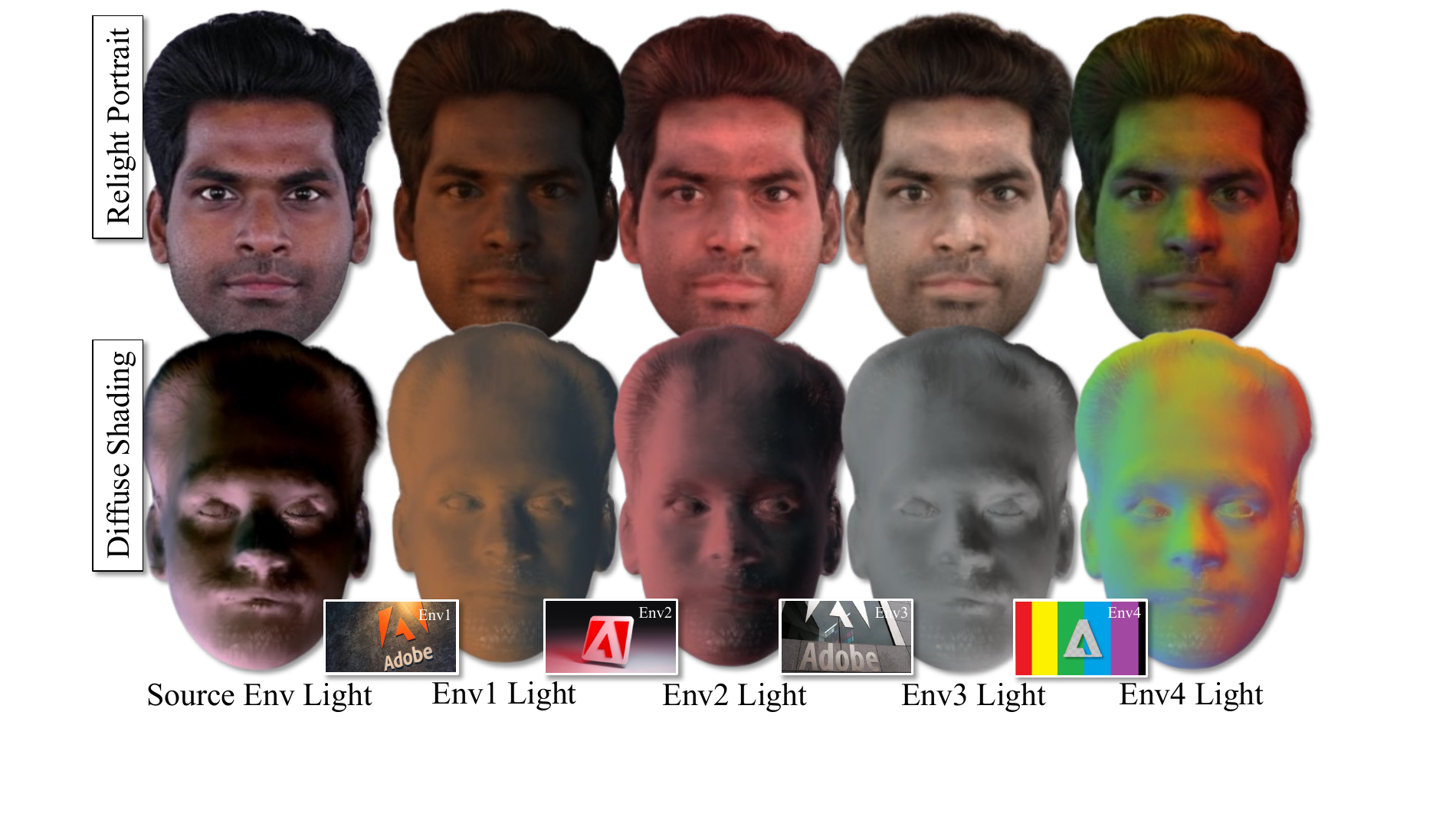}
  \caption{The visualize of relightable head reconstruction. From left to right, we set the source light source and the ambient light from different backgrounds to relight the face. Please pay attention to the colors reflected on the facial and geometry surface, which from the SH parameters re-rendering via the 3D Gaussian field. The artifacts around geometric and face due to the lack of 3D representation from single view setting. Natural face$\copyright$\textit{Wojciech Zielonka et al.} (CC BY), and background images$\copyright$\textit{Adobe} (CC BY).}
  \label{fig:app 1}
\end{figure}

Since the baseline methods do not support facial relighting, we exclude them from this discussion. We propose incorporating additional 3D Gaussian auxiliary properties during the warm-up phase, specifically with the spherical harmonics ($SH$) containing the lighting features. Furthermore, the specular reflections are adaptively decomposed during 3D Gaussian Splatting training. The accurate geometry and lighting decomposition allow for relighting by adjusting the pre-set light and light direction. Different from the previous methods~\cite{hou2024compose, qiu2024relitalk,cai2024real,lin2024edgerelight360}, our relighting training can be completed in just a few minutes, without the need for long-term training. However, constrained by the limited illumination information available in monocular video, our method cannot match the performance of approaches utilizing large datasets. Nonetheless, this represents an exploration of downstream applications and marks the first introduction of the concept of relighting in the monocular 3D Gaussian Splatting (3DGS) head avatar as far as we know. As shown in Figure~\ref{fig:app 1}, we present the relit appearance and geometry, the rendered face and diffuse shading are reflected in corresponding color of background images. The spherical harmonics are estimated from the provided background image, and the global illumination is derived by averaging the estimated results. We acknowledge that compare to some advanced relighting methods~\cite{he2024diffrelight,li2024uravatar,yoon2024generative}, our approach still has limitations in this application. However, it should be evaluated in the efficiency and monocular input setting.

\subsection{Facial Toonification Application}

\begin{figure}[t]
  \centering
  \includegraphics[width=1.\linewidth]{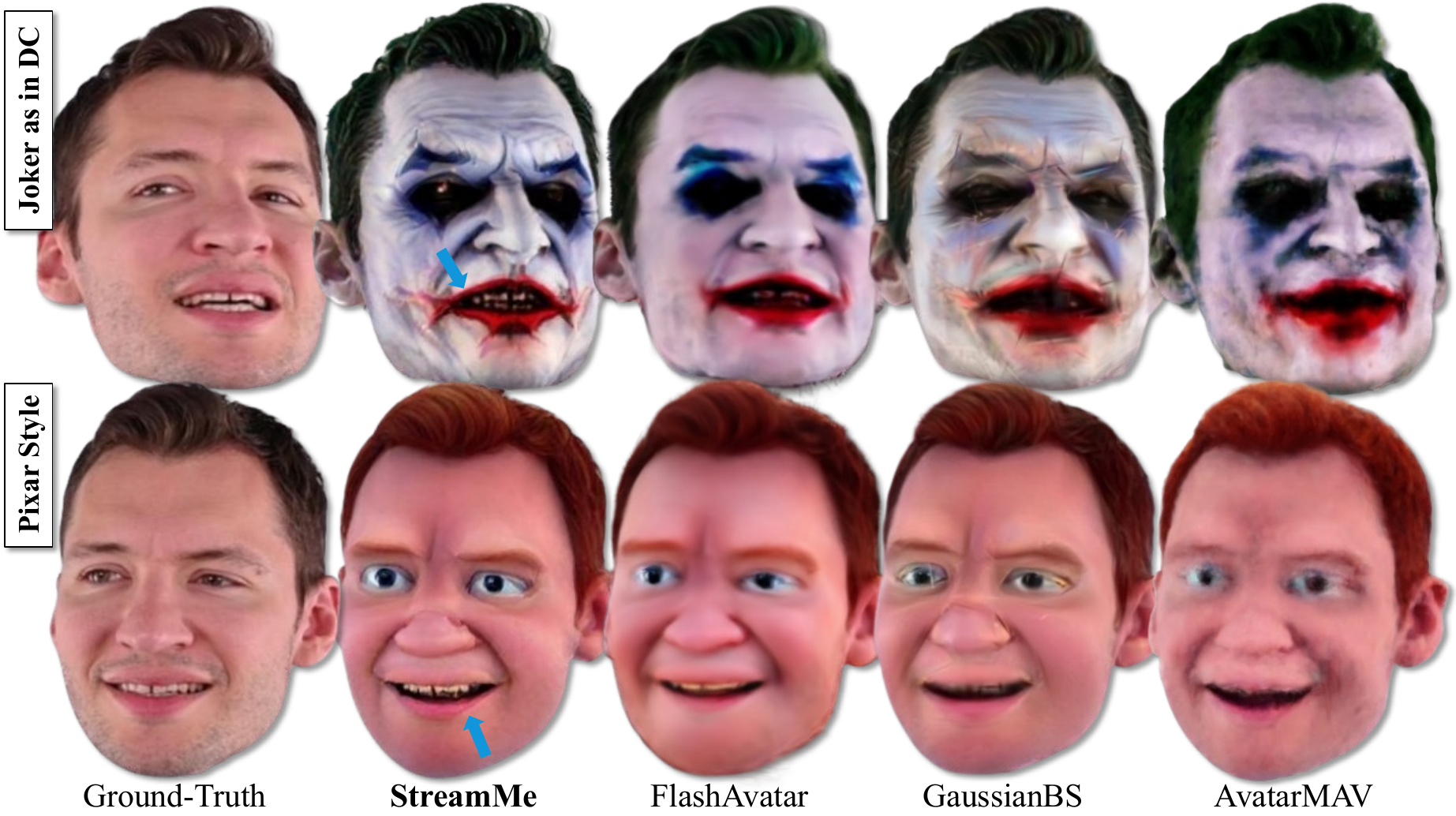}
  \caption{The visualize of toonification head reconstruction. We apply the TextToon~\cite{song2024texttoon} with two different prompts as \textit{"Joker in DC"} and \textit{"Pixar Style"} for the reconstructed appearance. We perform the perceptual ccomparison of toonification capabilities against the baseline methods. Blue arrows are used to highlight regions for attention. It can be found that our method achieves better toonification ability on details than the baselines with different styles. Natural face$\copyright$\textit{Wojciech Zielonka et al.} (CC BY).}
  \label{fig:app 2}
\end{figure}

We follow TextToon~\cite{song2024texttoon} and PortraitGen~\cite{gao2024portrait} to implement toonification (or stylization) application. These methods are based on stable diffusion~\cite{rombach2022high} for adaptive editing of the rendered images. It is worth noting that baseline methods are also capable of achieving adaptive editing. To provide a more comprehensive evaluation, we include editing with these methods for comparison. Specifically, we provide the Text2Image~\cite{brooks2023instructpix2pix} module with same setting (denoise steps, editing strength, guidance scale and boundary ratio~\textit{e.t.c.}) for baseline methods. The comparison results are shown in the Figure~\ref{fig:app 2}. As highlighted by the blue arrows, our method achieves richer details in texture editing compared to the baseline methods. 

\begin{figure}[h]
  \centering
  \includegraphics[width=1.\linewidth]{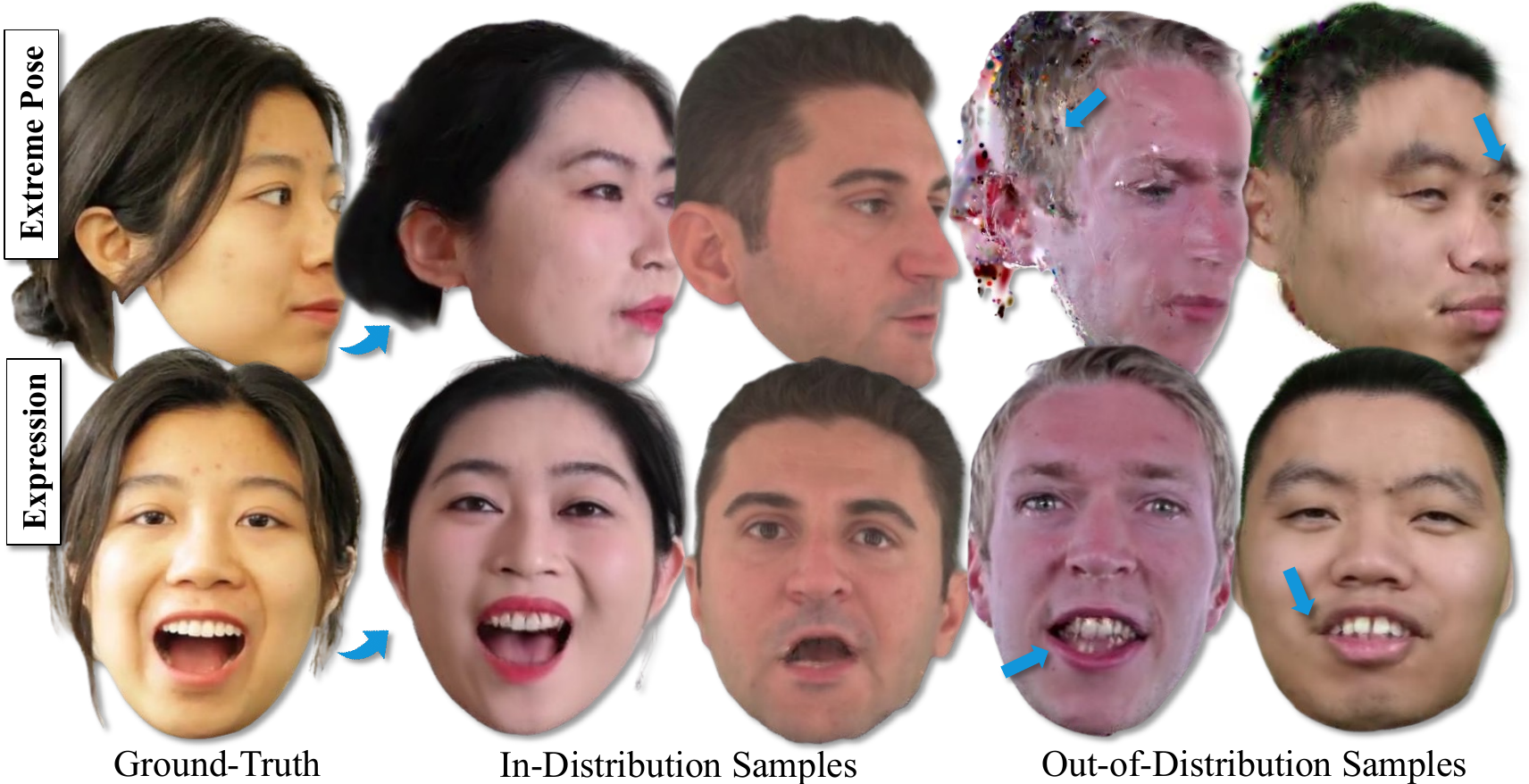}
  \caption{The visualize of limitation of monocular head reconstruction via our method. We present the results from cross-reenactment, as well as reconstruction outcomes for extreme poses and facial expressions. The in-distribution indicates that the training set includes similar poses and facial expressions, while out-of-distribution denotes the opposite. We use the blue arrows to highlight the artifacts. Natural face and $\copyright$\textit{Xuan Gao et al.} (CC BY).}
  \label{fig:limiation 1}
\end{figure}

\section{Limitations} The primary limitation of our method lies in its reliance on the single-view setting and the absence of a complete 3D facial structure, which hinders accurate reconstruction or learning of the full facial shape within the 3D Gaussian explicit space. This limitation often leads to rendering artifacts during head motion under extreme postures, as shown in first row in Figure~\ref{fig:limiation 1}. Second, the diversity of rendered appearances and facial expressions depends heavily on the training data, as in second row in Figure~\ref{fig:limiation 1}. Attempting to reconstruct appearances outside the distribution of the training data often yields disappointing results. It is worth emphasizing that this is not a limitation of our method specifically, but rather an inherent limitation of the single view setting, almost all related methods~\cite{xu2023avatarmav,wang2023styleavatar,song2024texttoon,kim2018deep,zheng2023pointavatar,chen2024monogaussianavatar} declare that.

\section{Discussion and Conclusion}
\label{Discussion and Conclusion}
We present StreamME, a on-the-fly head avatar training (or reconstruction) method from monocular live stream. It analyzes the number of point clouds in the 3D Gaussian field, evaluates the points distribution to achieve a more optimal arrangement over explicit face geometry, and removes points with lower contributions on scale and opacity to reduce computational complexity. The designs help our method achieves a diversity of facial expressions solely through geometry without dependence on the learnable MLPs, which significantly improves training speed. Furthermore, a series of applications (\textit{e.g.} toonification and relighting) have been developed based on our method, which bringing more exploration directions in the future.


\bibliographystyle{ACM-Reference-Format}
\bibliography{sample-bibliography}


\begin{thebibliography}{60}


\ifx \showCODEN    \undefined \def \showCODEN     #1{\unskip}     \fi
\ifx \showDOI      \undefined \def \showDOI       #1{#1}\fi
\ifx \showISBNx    \undefined \def \showISBNx     #1{\unskip}     \fi
\ifx \showISBNxiii \undefined \def \showISBNxiii  #1{\unskip}     \fi
\ifx \showISSN     \undefined \def \showISSN      #1{\unskip}     \fi
\ifx \showLCCN     \undefined \def \showLCCN      #1{\unskip}     \fi
\ifx \shownote     \undefined \def \shownote      #1{#1}          \fi
\ifx \showarticletitle \undefined \def \showarticletitle #1{#1}   \fi
\ifx \showURL      \undefined \def \showURL       {\relax}        \fi
\providecommand\bibfield[2]{#2}
\providecommand\bibinfo[2]{#2}
\providecommand\natexlab[1]{#1}
\providecommand\showeprint[2][]{arXiv:#2}

\bibitem[MIC(2022)]%
        {MICA:ECCV2022}
 \bibinfo{year}{2022}\natexlab{}.
\newblock \bibinfo{booktitle}{\emph{Towards Metrical Reconstruction of Human Faces}}.
\newblock


\bibitem[Bae et~al\mbox{.}(2024)]%
        {bae2024per}
\bibfield{author}{\bibinfo{person}{Jeongmin Bae}, \bibinfo{person}{Seoha Kim}, \bibinfo{person}{Youngsik Yun}, \bibinfo{person}{Hahyun Lee}, \bibinfo{person}{Gun Bang}, {and} \bibinfo{person}{Youngjung Uh}.} \bibinfo{year}{2024}\natexlab{}.
\newblock \showarticletitle{Per-Gaussian Embedding-Based Deformation for Deformable 3D Gaussian Splatting}.
\newblock \bibinfo{journal}{\emph{arXiv preprint arXiv:2404.03613}} (\bibinfo{year}{2024}).
\newblock


\bibitem[Bengio et~al\mbox{.}(2013)]%
        {bengio2013estimating}
\bibfield{author}{\bibinfo{person}{Yoshua Bengio}, \bibinfo{person}{Nicholas L{\'e}onard}, {and} \bibinfo{person}{Aaron Courville}.} \bibinfo{year}{2013}\natexlab{}.
\newblock \showarticletitle{Estimating or propagating gradients through stochastic neurons for conditional computation}.
\newblock \bibinfo{journal}{\emph{arXiv preprint arXiv:1308.3432}} (\bibinfo{year}{2013}).
\newblock


\bibitem[Brooks et~al\mbox{.}(2023)]%
        {brooks2023instructpix2pix}
\bibfield{author}{\bibinfo{person}{Tim Brooks}, \bibinfo{person}{Aleksander Holynski}, {and} \bibinfo{person}{Alexei~A Efros}.} \bibinfo{year}{2023}\natexlab{}.
\newblock \showarticletitle{Instructpix2pix: Learning to follow image editing instructions}. In \bibinfo{booktitle}{\emph{Proceedings of the IEEE/CVF Conference on Computer Vision and Pattern Recognition}}. \bibinfo{pages}{18392--18402}.
\newblock


\bibitem[Cai et~al\mbox{.}(2024)]%
        {cai2024real}
\bibfield{author}{\bibinfo{person}{Ziqi Cai}, \bibinfo{person}{Kaiwen Jiang}, \bibinfo{person}{Shu-Yu Chen}, \bibinfo{person}{Yu-Kun Lai}, \bibinfo{person}{Hongbo Fu}, \bibinfo{person}{Boxin Shi}, {and} \bibinfo{person}{Lin Gao}.} \bibinfo{year}{2024}\natexlab{}.
\newblock \showarticletitle{Real-time 3D-aware portrait video relighting}. In \bibinfo{booktitle}{\emph{Proceedings of the IEEE/CVF Conference on Computer Vision and Pattern Recognition}}. \bibinfo{pages}{6221--6231}.
\newblock


\bibitem[Cao et~al\mbox{.}(2022)]%
        {cao2022authentic}
\bibfield{author}{\bibinfo{person}{Chen Cao}, \bibinfo{person}{Tomas Simon}, \bibinfo{person}{Jin~Kyu Kim}, \bibinfo{person}{Gabe Schwartz}, \bibinfo{person}{Michael Zollhoefer}, \bibinfo{person}{Shun-Suke Saito}, \bibinfo{person}{Stephen Lombardi}, \bibinfo{person}{Shih-En Wei}, \bibinfo{person}{Danielle Belko}, \bibinfo{person}{Shoou-I Yu}, {et~al\mbox{.}}} \bibinfo{year}{2022}\natexlab{}.
\newblock \showarticletitle{Authentic volumetric avatars from a phone scan}.
\newblock \bibinfo{journal}{\emph{ACM Transactions on Graphics (TOG)}} \bibinfo{volume}{41}, \bibinfo{number}{4} (\bibinfo{year}{2022}), \bibinfo{pages}{1--19}.
\newblock


\bibitem[Chan et~al\mbox{.}(2022)]%
        {chan2022efficient}
\bibfield{author}{\bibinfo{person}{Eric~R Chan}, \bibinfo{person}{Connor~Z Lin}, \bibinfo{person}{Matthew~A Chan}, \bibinfo{person}{Koki Nagano}, \bibinfo{person}{Boxiao Pan}, \bibinfo{person}{Shalini De~Mello}, \bibinfo{person}{Orazio Gallo}, \bibinfo{person}{Leonidas~J Guibas}, \bibinfo{person}{Jonathan Tremblay}, \bibinfo{person}{Sameh Khamis}, {et~al\mbox{.}}} \bibinfo{year}{2022}\natexlab{}.
\newblock \showarticletitle{Efficient geometry-aware 3d generative adversarial networks}. In \bibinfo{booktitle}{\emph{Proceedings of the IEEE/CVF conference on computer vision and pattern recognition}}. \bibinfo{pages}{16123--16133}.
\newblock


\bibitem[Chen et~al\mbox{.}(2024)]%
        {chen2024monogaussianavatar}
\bibfield{author}{\bibinfo{person}{Yufan Chen}, \bibinfo{person}{Lizhen Wang}, \bibinfo{person}{Qijing Li}, \bibinfo{person}{Hongjiang Xiao}, \bibinfo{person}{Shengping Zhang}, \bibinfo{person}{Hongxun Yao}, {and} \bibinfo{person}{Yebin Liu}.} \bibinfo{year}{2024}\natexlab{}.
\newblock \showarticletitle{Monogaussianavatar: Monocular gaussian point-based head avatar}. In \bibinfo{booktitle}{\emph{ACM SIGGRAPH 2024 Conference Papers}}. \bibinfo{pages}{1--9}.
\newblock


\bibitem[Dhamo et~al\mbox{.}(2024)]%
        {dhamo2024headgas}
\bibfield{author}{\bibinfo{person}{Helisa Dhamo}, \bibinfo{person}{Yinyu Nie}, \bibinfo{person}{Arthur Moreau}, \bibinfo{person}{Jifei Song}, \bibinfo{person}{Richard Shaw}, \bibinfo{person}{Yiren Zhou}, {and} \bibinfo{person}{Eduardo P{\'e}rez-Pellitero}.} \bibinfo{year}{2024}\natexlab{}.
\newblock \showarticletitle{Headgas: Real-time animatable head avatars via 3d gaussian splatting}. In \bibinfo{booktitle}{\emph{European Conference on Computer Vision}}. Springer, \bibinfo{pages}{459--476}.
\newblock


\bibitem[Gafni et~al\mbox{.}(2021)]%
        {Gafni_2021_CVPR}
\bibfield{author}{\bibinfo{person}{Guy Gafni}, \bibinfo{person}{Justus Thies}, \bibinfo{person}{Michael Zollh{\"o}fer}, {and} \bibinfo{person}{Matthias Nie{\ss}ner}.} \bibinfo{year}{2021}\natexlab{}.
\newblock \showarticletitle{Dynamic Neural Radiance Fields for Monocular 4D Facial Avatar Reconstruction}. In \bibinfo{booktitle}{\emph{Proceedings of the IEEE/CVF Conference on Computer Vision and Pattern Recognition (CVPR)}}. \bibinfo{pages}{8649--8658}.
\newblock


\bibitem[Gao et~al\mbox{.}(2024)]%
        {gao2024portrait}
\bibfield{author}{\bibinfo{person}{Xuan Gao}, \bibinfo{person}{Haiyao Xiao}, \bibinfo{person}{Chenglai Zhong}, \bibinfo{person}{Shimin Hu}, \bibinfo{person}{Yudong Guo}, {and} \bibinfo{person}{Juyong Zhang}.} \bibinfo{year}{2024}\natexlab{}.
\newblock \showarticletitle{Portrait Video Editing Empowered by Multimodal Generative Priors}.
\newblock \bibinfo{journal}{\emph{arXiv preprint arXiv:2409.13591}} (\bibinfo{year}{2024}).
\newblock


\bibitem[Gao et~al\mbox{.}(2022)]%
        {Gao2022nerfblendshape}
\bibfield{author}{\bibinfo{person}{Xuan Gao}, \bibinfo{person}{Chenglai Zhong}, \bibinfo{person}{Jun Xiang}, \bibinfo{person}{Yang Hong}, \bibinfo{person}{Yudong Guo}, {and} \bibinfo{person}{Juyong Zhang}.} \bibinfo{year}{2022}\natexlab{}.
\newblock \showarticletitle{Reconstructing Personalized Semantic Facial NeRF Models From Monocular Video}.
\newblock \bibinfo{journal}{\emph{ACM Transactions on Graphics (Proceedings of SIGGRAPH Asia)}} \bibinfo{volume}{41}, \bibinfo{number}{6} (\bibinfo{year}{2022}).
\newblock
\urldef\tempurl%
\url{https://doi.org/10.1145/3550454.3555501}
\showDOI{\tempurl}


\bibitem[Giebenhain et~al\mbox{.}(2024)]%
        {giebenhain2024mononphm}
\bibfield{author}{\bibinfo{person}{Simon Giebenhain}, \bibinfo{person}{Tobias Kirschstein}, \bibinfo{person}{Markos Georgopoulos}, \bibinfo{person}{Martin R{\"{u}}nz}, \bibinfo{person}{Lourdes Agapito}, {and} \bibinfo{person}{Matthias Nie{\ss}ner}.} \bibinfo{year}{2024}\natexlab{}.
\newblock \showarticletitle{MonoNPHM: Dynamic Head Reconstruction from Monocular Videos}. In \bibinfo{booktitle}{\emph{Proc. IEEE Conf. on Computer Vision and Pattern Recognition (CVPR)}}.
\newblock


\bibitem[He et~al\mbox{.}(2024)]%
        {he2024diffrelight}
\bibfield{author}{\bibinfo{person}{Mingming He}, \bibinfo{person}{Pascal Clausen}, \bibinfo{person}{Ahmet~Levent Ta{\c{s}}el}, \bibinfo{person}{Li Ma}, \bibinfo{person}{Oliver Pilarski}, \bibinfo{person}{Wenqi Xian}, \bibinfo{person}{Laszlo Rikker}, \bibinfo{person}{Xueming Yu}, \bibinfo{person}{Ryan Burgert}, \bibinfo{person}{Ning Yu}, {et~al\mbox{.}}} \bibinfo{year}{2024}\natexlab{}.
\newblock \showarticletitle{DifFRelight: Diffusion-Based Facial Performance Relighting}.
\newblock \bibinfo{journal}{\emph{arXiv preprint arXiv:2410.08188}} (\bibinfo{year}{2024}).
\newblock


\bibitem[Hou et~al\mbox{.}(2024)]%
        {hou2024compose}
\bibfield{author}{\bibinfo{person}{Andrew Hou}, \bibinfo{person}{Zhixin Shu}, \bibinfo{person}{Xuaner Zhang}, \bibinfo{person}{He Zhang}, \bibinfo{person}{Yannick Hold-Geoffroy}, \bibinfo{person}{Jae~Shin Yoon}, {and} \bibinfo{person}{Xiaoming Liu}.} \bibinfo{year}{2024}\natexlab{}.
\newblock \showarticletitle{COMPOSE: Comprehensive Portrait Shadow Editing}.
\newblock \bibinfo{journal}{\emph{arXiv preprint arXiv:2408.13922}} (\bibinfo{year}{2024}).
\newblock


\bibitem[Kerbl et~al\mbox{.}(2023)]%
        {kerbl20233d}
\bibfield{author}{\bibinfo{person}{Bernhard Kerbl}, \bibinfo{person}{Georgios Kopanas}, \bibinfo{person}{Thomas Leimk{\"u}hler}, {and} \bibinfo{person}{George Drettakis}.} \bibinfo{year}{2023}\natexlab{}.
\newblock \showarticletitle{3D Gaussian Splatting for Real-Time Radiance Field Rendering}.
\newblock \bibinfo{journal}{\emph{ACM Transactions on Graphics}} \bibinfo{volume}{42}, \bibinfo{number}{4} (\bibinfo{year}{2023}).
\newblock


\bibitem[Kim et~al\mbox{.}(2018)]%
        {kim2018deep}
\bibfield{author}{\bibinfo{person}{Hyeongwoo Kim}, \bibinfo{person}{Pablo Garrido}, \bibinfo{person}{Ayush Tewari}, \bibinfo{person}{Weipeng Xu}, \bibinfo{person}{Justus Thies}, \bibinfo{person}{Matthias Niessner}, \bibinfo{person}{Patrick P{\'e}rez}, \bibinfo{person}{Christian Richardt}, \bibinfo{person}{Michael Zollh{\"o}fer}, {and} \bibinfo{person}{Christian Theobalt}.} \bibinfo{year}{2018}\natexlab{}.
\newblock \showarticletitle{Deep video portraits}.
\newblock \bibinfo{journal}{\emph{ACM transactions on graphics (TOG)}} \bibinfo{volume}{37}, \bibinfo{number}{4} (\bibinfo{year}{2018}), \bibinfo{pages}{1--14}.
\newblock


\bibitem[Lee et~al\mbox{.}(2024)]%
        {lee2024compact}
\bibfield{author}{\bibinfo{person}{Joo~Chan Lee}, \bibinfo{person}{Daniel Rho}, \bibinfo{person}{Xiangyu Sun}, \bibinfo{person}{Jong~Hwan Ko}, {and} \bibinfo{person}{Eunbyung Park}.} \bibinfo{year}{2024}\natexlab{}.
\newblock \showarticletitle{Compact 3d gaussian representation for radiance field}. In \bibinfo{booktitle}{\emph{Proceedings of the IEEE/CVF Conference on Computer Vision and Pattern Recognition}}. \bibinfo{pages}{21719--21728}.
\newblock


\bibitem[Li et~al\mbox{.}(2024)]%
        {li2024uravatar}
\bibfield{author}{\bibinfo{person}{Junxuan Li}, \bibinfo{person}{Chen Cao}, \bibinfo{person}{Gabriel Schwartz}, \bibinfo{person}{Rawal Khirodkar}, \bibinfo{person}{Christian Richardt}, \bibinfo{person}{Tomas Simon}, \bibinfo{person}{Yaser Sheikh}, {and} \bibinfo{person}{Shunsuke Saito}.} \bibinfo{year}{2024}\natexlab{}.
\newblock \showarticletitle{URAvatar: Universal Relightable Gaussian Codec Avatars}.
\newblock \bibinfo{journal}{\emph{arXiv preprint arXiv:2410.24223}} (\bibinfo{year}{2024}).
\newblock


\bibitem[Li et~al\mbox{.}(2023)]%
        {li2023spacetime}
\bibfield{author}{\bibinfo{person}{Zhan Li}, \bibinfo{person}{Zhang Chen}, \bibinfo{person}{Zhong Li}, {and} \bibinfo{person}{Yi Xu}.} \bibinfo{year}{2023}\natexlab{}.
\newblock \showarticletitle{Spacetime Gaussian Feature Splatting for Real-Time Dynamic View Synthesis}.
\newblock \bibinfo{journal}{\emph{arXiv preprint arXiv:2312.16812}} (\bibinfo{year}{2023}).
\newblock


\bibitem[Lin et~al\mbox{.}(2024)]%
        {lin2024edgerelight360}
\bibfield{author}{\bibinfo{person}{Min-Hui Lin}, \bibinfo{person}{Mahesh Reddy}, \bibinfo{person}{Guillaume Berger}, \bibinfo{person}{Michel Sarkis}, \bibinfo{person}{Fatih Porikli}, {and} \bibinfo{person}{Ning Bi}.} \bibinfo{year}{2024}\natexlab{}.
\newblock \showarticletitle{EdgeRelight360: Text-Conditioned 360-Degree HDR Image Generation for Real-Time On-Device Video Portrait Relighting}. In \bibinfo{booktitle}{\emph{Proceedings of the IEEE/CVF Conference on Computer Vision and Pattern Recognition}}. \bibinfo{pages}{831--840}.
\newblock


\bibitem[Lin et~al\mbox{.}(2022)]%
        {lin2022robust}
\bibfield{author}{\bibinfo{person}{Shanchuan Lin}, \bibinfo{person}{Linjie Yang}, \bibinfo{person}{Imran Saleemi}, {and} \bibinfo{person}{Soumyadip Sengupta}.} \bibinfo{year}{2022}\natexlab{}.
\newblock \showarticletitle{Robust high-resolution video matting with temporal guidance}. In \bibinfo{booktitle}{\emph{Proceedings of the IEEE/CVF Winter Conference on Applications of Computer Vision}}. \bibinfo{pages}{238--247}.
\newblock


\bibitem[Liu et~al\mbox{.}(2022)]%
        {liu2022deepfacevideoediting}
\bibfield{author}{\bibinfo{person}{Feng-Lin Liu}, \bibinfo{person}{Shu-Yu Chen}, \bibinfo{person}{Yu-Kun Lai}, \bibinfo{person}{Chunpeng Li}, \bibinfo{person}{Yue-Ren Jiang}, \bibinfo{person}{Hongbo Fu}, {and} \bibinfo{person}{Lin Gao}.} \bibinfo{year}{2022}\natexlab{}.
\newblock \showarticletitle{Deepfacevideoediting: Sketch-based deep editing of face videos}.
\newblock \bibinfo{journal}{\emph{ACM Transactions on Graphics (TOG)}} \bibinfo{volume}{41}, \bibinfo{number}{4} (\bibinfo{year}{2022}), \bibinfo{pages}{1--16}.
\newblock


\bibitem[Liu et~al\mbox{.}(2023)]%
        {liu2023instance}
\bibfield{author}{\bibinfo{person}{Yichen Liu}, \bibinfo{person}{Benran Hu}, \bibinfo{person}{Junkai Huang}, \bibinfo{person}{Yu-Wing Tai}, {and} \bibinfo{person}{Chi-Keung Tang}.} \bibinfo{year}{2023}\natexlab{}.
\newblock \showarticletitle{Instance neural radiance field}. In \bibinfo{booktitle}{\emph{Proceedings of the IEEE/CVF International Conference on Computer Vision}}. \bibinfo{pages}{787--796}.
\newblock


\bibitem[Luiten et~al\mbox{.}(2024)]%
        {luiten2023dynamic}
\bibfield{author}{\bibinfo{person}{Jonathon Luiten}, \bibinfo{person}{Georgios Kopanas}, \bibinfo{person}{Bastian Leibe}, {and} \bibinfo{person}{Deva Ramanan}.} \bibinfo{year}{2024}\natexlab{}.
\newblock \showarticletitle{Dynamic 3D Gaussians: Tracking by Persistent Dynamic View Synthesis}. In \bibinfo{booktitle}{\emph{3DV}}.
\newblock


\bibitem[Ma et~al\mbox{.}(2024)]%
        {ma20243d}
\bibfield{author}{\bibinfo{person}{Shengjie Ma}, \bibinfo{person}{Yanlin Weng}, \bibinfo{person}{Tianjia Shao}, {and} \bibinfo{person}{Kun Zhou}.} \bibinfo{year}{2024}\natexlab{}.
\newblock \showarticletitle{3D Gaussian Blendshapes for Head Avatar Animation}.
\newblock \bibinfo{journal}{\emph{arXiv preprint arXiv:2404.19398}} (\bibinfo{year}{2024}).
\newblock


\bibitem[Mihajlovic et~al\mbox{.}(2022)]%
        {mihajlovic2022keypointnerf}
\bibfield{author}{\bibinfo{person}{Marko Mihajlovic}, \bibinfo{person}{Aayush Bansal}, \bibinfo{person}{Michael Zollhoefer}, \bibinfo{person}{Siyu Tang}, {and} \bibinfo{person}{Shunsuke Saito}.} \bibinfo{year}{2022}\natexlab{}.
\newblock \showarticletitle{KeypointNeRF: Generalizing image-based volumetric avatars using relative spatial encoding of keypoints}. In \bibinfo{booktitle}{\emph{European conference on computer vision}}. Springer, \bibinfo{pages}{179--197}.
\newblock


\bibitem[Mildenhall et~al\mbox{.}(2020)]%
        {mildenhall2020nerf}
\bibfield{author}{\bibinfo{person}{Ben Mildenhall}, \bibinfo{person}{Pratul~P Srinivasan}, \bibinfo{person}{Matthew Tancik}, \bibinfo{person}{Jonathan~T Barron}, \bibinfo{person}{Ravi Ramamoorthi}, {and} \bibinfo{person}{Ren Ng}.} \bibinfo{year}{2020}\natexlab{}.
\newblock \showarticletitle{NeRF: Representing Scenes as Neural Radiance Fields for View Synthesis}. In \bibinfo{booktitle}{\emph{Computer Vision--ECCV 2020: 16th European Conference, Glasgow, UK, August 23--28, 2020, Proceedings, Part I}}. \bibinfo{pages}{405--421}.
\newblock


\bibitem[M{\"u}ller et~al\mbox{.}(2022)]%
        {muller2022instant}
\bibfield{author}{\bibinfo{person}{Thomas M{\"u}ller}, \bibinfo{person}{Alex Evans}, \bibinfo{person}{Christoph Schied}, {and} \bibinfo{person}{Alexander Keller}.} \bibinfo{year}{2022}\natexlab{}.
\newblock \showarticletitle{Instant neural graphics primitives with a multiresolution hash encoding}.
\newblock \bibinfo{journal}{\emph{arXiv preprint arXiv:2201.05989}} (\bibinfo{year}{2022}).
\newblock


\bibitem[Qian(2024)]%
        {qian2024versatile}
\bibfield{author}{\bibinfo{person}{Shenhan Qian}.} \bibinfo{year}{2024}\natexlab{}.
\newblock \showarticletitle{Versatile Head Alignment with Adaptive Appearance Priors}.
\newblock  (\bibinfo{date}{September} \bibinfo{year}{2024}).
\newblock
\urldef\tempurl%
\url{https://github.com/ShenhanQian/VHAP}
\showURL{%
\tempurl}


\bibitem[Qian et~al\mbox{.}(2023)]%
        {qian2023gaussianavatars}
\bibfield{author}{\bibinfo{person}{Shenhan Qian}, \bibinfo{person}{Tobias Kirschstein}, \bibinfo{person}{Liam Schoneveld}, \bibinfo{person}{Davide Davoli}, \bibinfo{person}{Simon Giebenhain}, {and} \bibinfo{person}{Matthias Nie{\ss}ner}.} \bibinfo{year}{2023}\natexlab{}.
\newblock \showarticletitle{GaussianAvatars: Photorealistic Head Avatars with Rigged 3D Gaussians}.
\newblock \bibinfo{journal}{\emph{arXiv preprint arXiv:2312.02069}} (\bibinfo{year}{2023}).
\newblock


\bibitem[Qian et~al\mbox{.}(2024)]%
        {qian20243dgs}
\bibfield{author}{\bibinfo{person}{Zhiyin Qian}, \bibinfo{person}{Shaofei Wang}, \bibinfo{person}{Marko Mihajlovic}, \bibinfo{person}{Andreas Geiger}, {and} \bibinfo{person}{Siyu Tang}.} \bibinfo{year}{2024}\natexlab{}.
\newblock \showarticletitle{3dgs-avatar: Animatable avatars via deformable 3d gaussian splatting}. In \bibinfo{booktitle}{\emph{Proceedings of the IEEE/CVF Conference on Computer Vision and Pattern Recognition}}. \bibinfo{pages}{5020--5030}.
\newblock


\bibitem[Qiu et~al\mbox{.}(2024)]%
        {qiu2024relitalk}
\bibfield{author}{\bibinfo{person}{Haonan Qiu}, \bibinfo{person}{Zhaoxi Chen}, \bibinfo{person}{Yuming Jiang}, \bibinfo{person}{Hang Zhou}, \bibinfo{person}{Xiangyu Fan}, \bibinfo{person}{Lei Yang}, \bibinfo{person}{Wayne Wu}, {and} \bibinfo{person}{Ziwei Liu}.} \bibinfo{year}{2024}\natexlab{}.
\newblock \showarticletitle{Relitalk: Relightable talking portrait generation from a single video}.
\newblock \bibinfo{journal}{\emph{International Journal of Computer Vision}} (\bibinfo{year}{2024}), \bibinfo{pages}{1--16}.
\newblock


\bibitem[Rombach et~al\mbox{.}(2022)]%
        {rombach2022high}
\bibfield{author}{\bibinfo{person}{Robin Rombach}, \bibinfo{person}{Andreas Blattmann}, \bibinfo{person}{Dominik Lorenz}, \bibinfo{person}{Patrick Esser}, {and} \bibinfo{person}{Bj{\"o}rn Ommer}.} \bibinfo{year}{2022}\natexlab{}.
\newblock \showarticletitle{High-resolution image synthesis with latent diffusion models}. In \bibinfo{booktitle}{\emph{Proceedings of the IEEE/CVF conference on computer vision and pattern recognition}}. \bibinfo{pages}{10684--10695}.
\newblock


\bibitem[Shao et~al\mbox{.}(2024)]%
        {shao2024splattingavatar}
\bibfield{author}{\bibinfo{person}{Zhijing Shao}, \bibinfo{person}{Zhaolong Wang}, \bibinfo{person}{Zhuang Li}, \bibinfo{person}{Duotun Wang}, \bibinfo{person}{Xiangru Lin}, \bibinfo{person}{Yu Zhang}, \bibinfo{person}{Mingming Fan}, {and} \bibinfo{person}{Zeyu Wang}.} \bibinfo{year}{2024}\natexlab{}.
\newblock \showarticletitle{Splattingavatar: Realistic real-time human avatars with mesh-embedded gaussian splatting}.
\newblock \bibinfo{journal}{\emph{arXiv preprint arXiv:2403.05087}} (\bibinfo{year}{2024}).
\newblock


\bibitem[Shen et~al\mbox{.}(2021)]%
        {shen2021deep}
\bibfield{author}{\bibinfo{person}{Tianchang Shen}, \bibinfo{person}{Jun Gao}, \bibinfo{person}{Kangxue Yin}, \bibinfo{person}{Ming-Yu Liu}, {and} \bibinfo{person}{Sanja Fidler}.} \bibinfo{year}{2021}\natexlab{}.
\newblock \showarticletitle{Deep marching tetrahedra: a hybrid representation for high-resolution 3d shape synthesis}.
\newblock \bibinfo{journal}{\emph{Advances in Neural Information Processing Systems}}  \bibinfo{volume}{34} (\bibinfo{year}{2021}), \bibinfo{pages}{6087--6101}.
\newblock


\bibitem[Song et~al\mbox{.}(2024a)]%
        {song2024texttoon}
\bibfield{author}{\bibinfo{person}{Luchuan Song}, \bibinfo{person}{Lele Chen}, \bibinfo{person}{Celong Liu}, \bibinfo{person}{Pinxin Liu}, {and} \bibinfo{person}{Chenliang Xu}.} \bibinfo{year}{2024}\natexlab{a}.
\newblock \showarticletitle{TextToon: Real-Time Text Toonify Head Avatar from Single Video}.
\newblock \bibinfo{journal}{\emph{arXiv preprint arXiv:2410.07160}} (\bibinfo{year}{2024}).
\newblock


\bibitem[Song et~al\mbox{.}(2021a)]%
        {song2021tacr}
\bibfield{author}{\bibinfo{person}{Luchuan Song}, \bibinfo{person}{Bin Liu}, \bibinfo{person}{Guojun Yin}, \bibinfo{person}{Xiaoyi Dong}, \bibinfo{person}{Yufei Zhang}, {and} \bibinfo{person}{Jia-Xuan Bai}.} \bibinfo{year}{2021}\natexlab{a}.
\newblock \showarticletitle{Tacr-net: editing on deep video and voice portraits}. In \bibinfo{booktitle}{\emph{Proceedings of the 29th ACM International Conference on Multimedia}}. \bibinfo{pages}{478--486}.
\newblock


\bibitem[Song et~al\mbox{.}(2024b)]%
        {song2024tri}
\bibfield{author}{\bibinfo{person}{Luchuan Song}, \bibinfo{person}{Pinxin Liu}, \bibinfo{person}{Lele Chen}, \bibinfo{person}{Guojun Yin}, {and} \bibinfo{person}{Chenliang Xu}.} \bibinfo{year}{2024}\natexlab{b}.
\newblock \showarticletitle{Tri 2-plane: Thinking Head Avatar via Feature Pyramid}. In \bibinfo{booktitle}{\emph{European Conference on Computer Vision}}. Springer, \bibinfo{pages}{1--20}.
\newblock


\bibitem[Song et~al\mbox{.}(2023)]%
        {song2023emotional}
\bibfield{author}{\bibinfo{person}{Luchuan Song}, \bibinfo{person}{Guojun Yin}, \bibinfo{person}{Zhenchao Jin}, \bibinfo{person}{Xiaoyi Dong}, {and} \bibinfo{person}{Chenliang Xu}.} \bibinfo{year}{2023}\natexlab{}.
\newblock \showarticletitle{Emotional listener portrait: Neural listener head generation with emotion}. In \bibinfo{booktitle}{\emph{Proceedings of the IEEE/CVF International Conference on Computer Vision}}. \bibinfo{pages}{20839--20849}.
\newblock


\bibitem[Song et~al\mbox{.}(2021b)]%
        {song2021fsft}
\bibfield{author}{\bibinfo{person}{Luchuan Song}, \bibinfo{person}{Guojun Yin}, \bibinfo{person}{Bin Liu}, \bibinfo{person}{Yuhui Zhang}, {and} \bibinfo{person}{Nenghai Yu}.} \bibinfo{year}{2021}\natexlab{b}.
\newblock \showarticletitle{Fsft-net: face transfer video generation with few-shot views}. In \bibinfo{booktitle}{\emph{2021 IEEE international conference on image processing (ICIP)}}. IEEE, \bibinfo{pages}{3582--3586}.
\newblock


\bibitem[Thies et~al\mbox{.}(2016)]%
        {thies2016face2face}
\bibfield{author}{\bibinfo{person}{Justus Thies}, \bibinfo{person}{Michael Zollhofer}, \bibinfo{person}{Marc Stamminger}, \bibinfo{person}{Christian Theobalt}, {and} \bibinfo{person}{Matthias Nie{\ss}ner}.} \bibinfo{year}{2016}\natexlab{}.
\newblock \showarticletitle{Face2face: Real-time face capture and reenactment of rgb videos}. In \bibinfo{booktitle}{\emph{Proceedings of the IEEE conference on computer vision and pattern recognition}}. \bibinfo{pages}{2387--2395}.
\newblock


\bibitem[Tu et~al\mbox{.}(2024)]%
        {tu2024tele}
\bibfield{author}{\bibinfo{person}{Hanzhang Tu}, \bibinfo{person}{Ruizhi Shao}, \bibinfo{person}{Xue Dong}, \bibinfo{person}{Shunyuan Zheng}, \bibinfo{person}{Hao Zhang}, \bibinfo{person}{Lili Chen}, \bibinfo{person}{Meili Wang}, \bibinfo{person}{Wenyu Li}, \bibinfo{person}{Siyan Ma}, \bibinfo{person}{Shengping Zhang}, {et~al\mbox{.}}} \bibinfo{year}{2024}\natexlab{}.
\newblock \showarticletitle{Tele-Aloha: A Low-budget and High-authenticity Telepresence System Using Sparse RGB Cameras}.
\newblock \bibinfo{journal}{\emph{arXiv preprint arXiv:2405.14866}} (\bibinfo{year}{2024}).
\newblock


\bibitem[Van Den~Oord et~al\mbox{.}(2017)]%
        {van2017neural}
\bibfield{author}{\bibinfo{person}{Aaron Van Den~Oord}, \bibinfo{person}{Oriol Vinyals}, {et~al\mbox{.}}} \bibinfo{year}{2017}\natexlab{}.
\newblock \showarticletitle{Neural discrete representation learning}.
\newblock \bibinfo{journal}{\emph{Advances in neural information processing systems}}  \bibinfo{volume}{30} (\bibinfo{year}{2017}).
\newblock


\bibitem[Wang et~al\mbox{.}(2023)]%
        {wang2023styleavatar}
\bibfield{author}{\bibinfo{person}{Lizhen Wang}, \bibinfo{person}{Xiaochen Zhao}, \bibinfo{person}{Jingxiang Sun}, \bibinfo{person}{Yuxiang Zhang}, \bibinfo{person}{Hongwen Zhang}, \bibinfo{person}{Tao Yu}, {and} \bibinfo{person}{Yebin Liu}.} \bibinfo{year}{2023}\natexlab{}.
\newblock \showarticletitle{StyleAvatar: Real-time Photo-realistic Portrait Avatar from a Single Video}.
\newblock \bibinfo{journal}{\emph{arXiv preprint arXiv:2305.00942}} (\bibinfo{year}{2023}).
\newblock


\bibitem[Wang et~al\mbox{.}(2022)]%
        {wang2022hvh}
\bibfield{author}{\bibinfo{person}{Ziyan Wang}, \bibinfo{person}{Giljoo Nam}, \bibinfo{person}{Tuur Stuyck}, \bibinfo{person}{Stephen Lombardi}, \bibinfo{person}{Michael Zollh{\"o}fer}, \bibinfo{person}{Jessica Hodgins}, {and} \bibinfo{person}{Christoph Lassner}.} \bibinfo{year}{2022}\natexlab{}.
\newblock \showarticletitle{Hvh: Learning a hybrid neural volumetric representation for dynamic hair performance capture}. In \bibinfo{booktitle}{\emph{Proceedings of the IEEE/CVF Conference on Computer Vision and Pattern Recognition}}. \bibinfo{pages}{6143--6154}.
\newblock


\bibitem[Xiang et~al\mbox{.}(2023)]%
        {xiang2023flashavatar}
\bibfield{author}{\bibinfo{person}{Jun Xiang}, \bibinfo{person}{Xuan Gao}, \bibinfo{person}{Yudong Guo}, {and} \bibinfo{person}{Juyong Zhang}.} \bibinfo{year}{2023}\natexlab{}.
\newblock \showarticletitle{FlashAvatar: High-Fidelity Digital Avatar Rendering at 300FPS}.
\newblock \bibinfo{journal}{\emph{arXiv preprint arXiv:2312.02214}} (\bibinfo{year}{2023}).
\newblock


\bibitem[Xu et~al\mbox{.}(2024)]%
        {xu2023gaussianheadavatar}
\bibfield{author}{\bibinfo{person}{Yuelang Xu}, \bibinfo{person}{Benwang Chen}, \bibinfo{person}{Zhe Li}, \bibinfo{person}{Hongwen Zhang}, \bibinfo{person}{Lizhen Wang}, \bibinfo{person}{Zerong Zheng}, {and} \bibinfo{person}{Yebin Liu}.} \bibinfo{year}{2024}\natexlab{}.
\newblock \showarticletitle{Gaussian Head Avatar: Ultra High-fidelity Head Avatar via Dynamic Gaussians}. In \bibinfo{booktitle}{\emph{Proceedings of the IEEE/CVF Conference on Computer Vision and Pattern Recognition (CVPR)}}.
\newblock


\bibitem[Xu et~al\mbox{.}(2023)]%
        {xu2023avatarmav}
\bibfield{author}{\bibinfo{person}{Yuelang Xu}, \bibinfo{person}{Lizhen Wang}, \bibinfo{person}{Xiaochen Zhao}, \bibinfo{person}{Hongwen Zhang}, {and} \bibinfo{person}{Yebin Liu}.} \bibinfo{year}{2023}\natexlab{}.
\newblock \showarticletitle{AvatarMAV: Fast 3D Head Avatar Reconstruction Using Motion-Aware Neural Voxels}. In \bibinfo{booktitle}{\emph{ACM SIGGRAPH 2023 Conference Proceedings}}.
\newblock


\bibitem[Yoon et~al\mbox{.}(2024)]%
        {yoon2024generative}
\bibfield{author}{\bibinfo{person}{Jae~Shin Yoon}, \bibinfo{person}{Zhixin Shu}, \bibinfo{person}{Mengwei Ren}, \bibinfo{person}{Xuaner Zhang}, \bibinfo{person}{Yannick Hold-Geoffroy}, \bibinfo{person}{Krishna~Kumar Singh}, {and} \bibinfo{person}{He Zhang}.} \bibinfo{year}{2024}\natexlab{}.
\newblock \showarticletitle{Generative Portrait Shadow Removal}.
\newblock \bibinfo{journal}{\emph{arXiv preprint arXiv:2410.05525}} (\bibinfo{year}{2024}).
\newblock


\bibitem[Zhang et~al\mbox{.}(2023)]%
        {zhang2023deformtoon3d}
\bibfield{author}{\bibinfo{person}{Junzhe Zhang}, \bibinfo{person}{Yushi Lan}, \bibinfo{person}{Shuai Yang}, \bibinfo{person}{Fangzhou Hong}, \bibinfo{person}{Quan Wang}, \bibinfo{person}{Chai~Kiat Yeo}, \bibinfo{person}{Ziwei Liu}, {and} \bibinfo{person}{Chen~Change Loy}.} \bibinfo{year}{2023}\natexlab{}.
\newblock \showarticletitle{Deformtoon3d: Deformable neural radiance fields for 3d toonification}. In \bibinfo{booktitle}{\emph{Proceedings of the IEEE/CVF International Conference on Computer Vision}}. \bibinfo{pages}{9144--9154}.
\newblock


\bibitem[Zhang et~al\mbox{.}(2018)]%
        {zhang2018perceptual}
\bibfield{author}{\bibinfo{person}{Richard Zhang}, \bibinfo{person}{Phillip Isola}, \bibinfo{person}{Alexei~A Efros}, \bibinfo{person}{Eli Shechtman}, {and} \bibinfo{person}{Oliver Wang}.} \bibinfo{year}{2018}\natexlab{}.
\newblock \showarticletitle{The Unreasonable Effectiveness of Deep Features as a Perceptual Metric}. In \bibinfo{booktitle}{\emph{CVPR}}.
\newblock


\bibitem[Zhang et~al\mbox{.}(2024a)]%
        {zhang2024discover}
\bibfield{author}{\bibinfo{person}{Zeliang Zhang}, \bibinfo{person}{Mingqian Feng}, \bibinfo{person}{Zhiheng Li}, {and} \bibinfo{person}{Chenliang Xu}.} \bibinfo{year}{2024}\natexlab{a}.
\newblock \showarticletitle{Discover and mitigate multiple biased subgroups in image classifiers}. In \bibinfo{booktitle}{\emph{Proceedings of the IEEE/CVF Conference on Computer Vision and Pattern Recognition}}. \bibinfo{pages}{10906--10915}.
\newblock


\bibitem[Zhang et~al\mbox{.}(2024b)]%
        {zhang2024bag}
\bibfield{author}{\bibinfo{person}{Zeliang Zhang}, \bibinfo{person}{Wei Yao}, {and} \bibinfo{person}{Xiaosen Wang}.} \bibinfo{year}{2024}\natexlab{b}.
\newblock \showarticletitle{Bag of tricks to boost adversarial transferability}.
\newblock \bibinfo{journal}{\emph{arXiv preprint arXiv:2401.08734}} (\bibinfo{year}{2024}).
\newblock


\bibitem[Zheng et~al\mbox{.}(2022)]%
        {zheng2022avatar}
\bibfield{author}{\bibinfo{person}{Yufeng Zheng}, \bibinfo{person}{Victoria~Fern{\'a}ndez Abrevaya}, \bibinfo{person}{Marcel~C B{\"u}hler}, \bibinfo{person}{Xu Chen}, \bibinfo{person}{Michael~J Black}, {and} \bibinfo{person}{Otmar Hilliges}.} \bibinfo{year}{2022}\natexlab{}.
\newblock \showarticletitle{Im avatar: Implicit morphable head avatars from videos}. In \bibinfo{booktitle}{\emph{Proceedings of the IEEE/CVF Conference on Computer Vision and Pattern Recognition}}. \bibinfo{pages}{13545--13555}.
\newblock


\bibitem[Zheng et~al\mbox{.}(2023)]%
        {zheng2023pointavatar}
\bibfield{author}{\bibinfo{person}{Yufeng Zheng}, \bibinfo{person}{Wang Yifan}, \bibinfo{person}{Gordon Wetzstein}, \bibinfo{person}{Michael~J Black}, {and} \bibinfo{person}{Otmar Hilliges}.} \bibinfo{year}{2023}\natexlab{}.
\newblock \showarticletitle{Pointavatar: Deformable point-based head avatars from videos}. In \bibinfo{booktitle}{\emph{Proceedings of the IEEE/CVF Conference on Computer Vision and Pattern Recognition}}. \bibinfo{pages}{21057--21067}.
\newblock


\bibitem[Zhu et~al\mbox{.}(2024)]%
        {zhu2024motiongs}
\bibfield{author}{\bibinfo{person}{Ruijie Zhu}, \bibinfo{person}{Yanzhe Liang}, \bibinfo{person}{Hanzhi Chang}, \bibinfo{person}{Jiacheng Deng}, \bibinfo{person}{Jiahao Lu}, \bibinfo{person}{Wenfei Yang}, \bibinfo{person}{Tianzhu Zhang}, {and} \bibinfo{person}{Yongdong Zhang}.} \bibinfo{year}{2024}\natexlab{}.
\newblock \showarticletitle{MotionGS: Exploring Explicit Motion Guidance for Deformable 3D Gaussian Splatting}.
\newblock \bibinfo{journal}{\emph{arXiv preprint arXiv:2410.07707}} (\bibinfo{year}{2024}).
\newblock


\bibitem[Zielonka et~al\mbox{.}(2024)]%
        {zielonka2024gaussian}
\bibfield{author}{\bibinfo{person}{Wojciech Zielonka}, \bibinfo{person}{Timo Bolkart}, \bibinfo{person}{Thabo Beeler}, {and} \bibinfo{person}{Justus Thies}.} \bibinfo{year}{2024}\natexlab{}.
\newblock \showarticletitle{Gaussian Eigen Models for Human Heads}.
\newblock \bibinfo{journal}{\emph{arXiv preprint arXiv:2407.04545}} (\bibinfo{year}{2024}).
\newblock


\bibitem[Zielonka et~al\mbox{.}(2022)]%
        {zielonka2022towards}
\bibfield{author}{\bibinfo{person}{Wojciech Zielonka}, \bibinfo{person}{Timo Bolkart}, {and} \bibinfo{person}{Justus Thies}.} \bibinfo{year}{2022}\natexlab{}.
\newblock \showarticletitle{Towards metrical reconstruction of human faces}. In \bibinfo{booktitle}{\emph{European Conference on Computer Vision}}. Springer, \bibinfo{pages}{250--269}.
\newblock


\bibitem[Zielonka et~al\mbox{.}(2023)]%
        {zielonka2023instant}
\bibfield{author}{\bibinfo{person}{Wojciech Zielonka}, \bibinfo{person}{Timo Bolkart}, {and} \bibinfo{person}{Justus Thies}.} \bibinfo{year}{2023}\natexlab{}.
\newblock \showarticletitle{Instant volumetric head avatars}. In \bibinfo{booktitle}{\emph{Proceedings of the IEEE/CVF Conference on Computer Vision and Pattern Recognition}}. \bibinfo{pages}{4574--4584}.
\newblock


\end{thebibliography}

\newpage

\end{document}